\documentclass[twocolumn,showpacs,prb,floatfix,amsmath,amssymb]{revtex4}

\usepackage{amssymb}
\usepackage{amsmath}
\usepackage{bm}
\usepackage{bbm}
\usepackage{graphicx}
\usepackage{float}

\DeclareMathOperator{\Real}{Re} 
\DeclareMathAlphabet{\mathpzc}{OT1}{pzc}{m}{it}

\newcommand{\transpose}{^{\scriptscriptstyle\mathrm T}}

\newcommand{\identity}{\mathbbm{1}}

\newcommand{\HD}[2]{\hat{\cal H}^{\mbox{\scriptsize D}}_{#1 #2}}
\newcommand{\HEx}[2]{\hat{\cal H}^{\mbox{\scriptsize Ex}}_{#1 #2}}

\newcommand{\HHFM}{\hat{\cal H}^{\mbox{\scriptsize HFM}}}
\newcommand{\rVec}{\vec{r}}
\newcommand{\Rvec}{\vec{R}}
\newcommand{\clusterCountW}[1]{\hat{W}^{[#1]}}
\newcommand{\clusterW}[1]{\hat{W}_{#1}'}
\newcommand{\clusterContrib}[1]{\langle \hat{W}_{#1}' \rangle}
\newcommand{\biClusterW}[2]{\hat{W}_{#1, #2}'}

\newcommand{\setW}[1]{\hat{W}_{#1}}
\newcommand{\bigO}[1]{\ensuremath{{\cal O}\left( #1 \right)}}

\begin{document}

\title{Wavefunction considerations for the central
spin decoherence problem in a nuclear spin bath}
\author{W. M. \surname{Witzel}$^{1,2}$, S. \surname{Das Sarma}$^{1}$}
\affiliation{$^{1}$Condensed Matter Theory Center,
  Department of Physics, University of Maryland, College Park, Maryland
  20742-4111, USA \\
$^{2}$Naval Research Laboratory, Washington, DC 20375, USA} 

\begin{abstract}
Decoherence of a localized electron spin in a solid state material 
(the ``central spin'' problem)
at low
temperature is believed to be dominated by interactions with
nuclear spins in
the lattice.  This decoherence is partially suppressed through the
application of a large magnetic field that splits the energy levels of
the electron spin and prevents depolarization.  However, the dephasing
decoherence resulting from a dynamical nuclear spin bath cannot be
removed in this way.  Fluctuations of the nuclear field lead to an
uncertainty of the electron's precessional frequency in a process
known as spectral diffusion.  This paper considers the effect of
the electron's wavefunction shape on spectral diffusion and
provides wavefunction dependent decoherence time formulas for a free
induction decay as well as spin echoes and concatenated dynamical decoupling
schemes for enhancing coherence.  We also discuss a dephasing of a qubit encoded
in singlet-triplet states of a double quantum dot.
A central theoretical result of this work is the development of a
continuum approximation for the spectral diffusion problem which we
have applied to GaAs and InAs materials specifically.
\end{abstract}
\maketitle

\section{Introduction}

Understanding quantum decoherence is a fundamental
subject of interest in modern physics.  In this work, we theoretically
study the issue of quantum decoherence for the problem of one
localized electron spin in a solid state nuclear spin environment,
where the electron spin eventually loses its quantum phase memory
(i.e., dephases) due to its interaction with the surrounding nuclear
spin bath.  
This is often called ``central spin'' decoherence in a spin
bath, with the localized electron spin being the central spin and the
surrounding nuclear spin environment being the spin bath.
This particular problem is important in the context of
quantum information processing and quantum computation using localized
electron spins as qubits, and as such, we concentrate on a few
systems of interest in solid state quantum computation architectures,
namely, Si:P donor electron spin qubits and GaAs and InAs quantum dot electron
spin qubits, all of which have considerable recent
experimental\cite{tyryshkin, abe, petta05, Delft, modeLock} and
theoretical\cite{desousa03b, Basel, witzelHahnShort, yaoHahn, deng, witzelHahnLong,
  witzelCPMG, yaoCDD, yaoCDDlong, witzelCDD} 
 interest.  The theory we develop is, however, applicable
to the general situation of the quantum dephasing of a single
localized electron spin in solids due to the environmental influence
of the slowly fluctuating nuclear spin bath consisting of many
millions of surrounding nuclear spins mutually flip-flopping due to
their magnetic dipolar coupling.

The issue of specific interest in this paper, as the title of this
paper suggests, is how the detailed form of the confinement for the
localized electron in the solid (e.g., the exponentially confined
hydrogenic confinement for the localized P donor electron state in Si
or the Gaussian-type simple harmonic oscillator confinement for the
localized electron in the GaAs/InAs quantum dot) could have qualitative
influence on its nuclear induced spin dephasing.  This subtle (but
potentially significant) dependence of electron spin
dephasing on the nature of the electron localization has recently been
emphasized in the discovery\cite{UDD} that a particular type of
dynamical decoupling (DD) sequence\cite{UhrigSequence} can be ideal in
restoring quantum coherence in the GaAs quantum dot system, but not
particularly effective in the Si:P system, which can be traced back to
the Gaussian versus the exponential wavefunction localization in the
two systems, leading to the validity or the lack thereof of a
particular time perturbation expansion as discussed in depth in this
work.  Thus, a detailed investigation of the effect of the localized
electron wavefunction on the nuclear induced electron spin dephasing
problem is both important and timely in view of the intense current
activity in fault-tolerant quantum computation using spin qubits in
semiconductors.

It is important in the context of studying electron spin decoherence
to distinguish among three different spin relaxation or decoherence
times, $T_1$, $T_2^{*}$, and $T_2$, which are discussed in the
literature.  (We should mention right at the outset that our work is
focused entirely on $T_2$, sometimes also denoted $T_M$ or spin memory
time.  $T_2$ is variously called spin decoherence
time, spin dephasing time, transverse spin relaxation time, spin-spin
relaxation time, and spin memory time in the literature.)  The spin
relaxation time $T_1$, also often called the longitudinal spin
relaxation time or the energy relaxation time, is connected with the
spin flip process which, in the presence of an externally applied
magnetic field (the case of interest to us in this work),
necessarily requires phonons (and spin-orbit coupling) to carry away
the electron spin Zeeman energy, which is 3 orders of magnitude
larger than the nuclear spin Zeeman energy.  This $T_1$-relaxation
process can be made arbitrarily long by lowering the lattice
temperature so that phonons are simply not available to provide the
energy conservation.  At the low ($\sim 100~\mbox{mK}$ or lower)
temperatures of interest to us in the quantum computing context, the
relevant $T_1$ times are very long ($T_1 > 100~\mbox{ms} \gg T_2$) and
are unimportant for our consideration.  The $T_2^{*}$ time is the
relevant decoherence time in the presence of substantial inhomogeneous
broadening as, for example, in ensemble measurements over many
electron spin qubits with varying (i.e., inhomogeneous) nuclear spin
environments.  In the context of single spin qubits, i.e. involving a
single electron spin, the $T_2^{*}$ decoherence sets in due to the
requisite time averaging which, due to ergodicity, becomes equivalent
to the spatial inhomogeneity of the varying nuclear spin environments
of many electron spins.  Thus, $T_2^{*}$ is measured either in a
measurement over an ensemble of localized spins with the associated
spatial averaging or in a time-averaged measurement for a single spin
over many runs.  A spin echo (or Hahn spin echo) measurement gets rid
of the inhomogeneous broadening and characterizes $T_2$, the pure dephasing
time of a single spin (typically for the systems of our interest
$T_2^{*} \lesssim T_2 / 1000$ and $T_2 < T_1 / 1000$), which is what we
theoretically study in this work.  A closely related, but by no means
identical, definition of $T_2$ comes from considering the free
induction decay (FID) of a single spin in a single-shot measurement
without involving either spatial averaging over many spin qubits or
temporal averaging over many runs.  Alternatively, FID is observed in
a homogeneous ensemble.
We will call such a FID dephasing
time $T_I~(\lesssim T_2)$ to distinguish it from the spin echo
dephasing time $T_2$.  

The above discussion of $T_1$, $T_2^*$, $T_2$, and $T_I$ illustrates
the considerable semantic danger of discussing ``spin decoherence''
because, depending on the context, the ``spin decoherence time'' for
the same system could vary by many orders of magnitude (i.e., $T_1 \gg
T_2 \gtrsim T_I \gg T_2^*$, etc.).  To avoid such confusion, we
emphasize that, in our opinion, the only sensible way of discussing
spin decoherence is by considering specific experimental contexts.
Our definition of $T_2$ is thus the decoherence time measured in a
Hahn spin echo experiment.  The only license we take with our
definition of $T_2$ is that we continue using $T_2$ as the notation
for spin decoherenc time even in situations where the spin coherence
has been extended far beyond the Hahn spin echo time by using multiple
pulse sequences [Carr-Purcell-Meiboom-Gill (CPMG), concatenated
  dynamical decoupling (CDD), etc.] much more complex than the single
$\pi$-pulse Hahn sequence.  For us, therefore, $T_2$ is the spin
decoherence time as measured in an echo-type pulse sequence
measurement, which could be a simple $\pi$-pulse spin echo or more
complex pulse sequences meant to prolong spin memory beyond the spin
echo refocusing.

Finally, we point out an additional important complication, often
erroneously neglected in the literature, associated with discussing
spin decoherence in terms of a single decoherence time parameter,
$T_{\mbox{\tiny Coh}}$
(e.g., $T_1$ or $T_2$ or $T_2^*$ or $T_I$, etc.).  Such a description
assumes, by definition, that the quantum memory (i.e., some precisely
defined quantum amplitude or probability) falls off in a simple
exponential manner with time, i.e., $\exp{(-t / T_{\mbox{\tiny Coh}})}$
or $\exp{(-[t / T_{\mbox{\tiny Coh}}]^n)}$, where $n$ is a constant, so
that a single decoherence time $T_{\mbox{\tiny Coh}}$ can completely
parametrize the nature of decoherence.  This is, however, not always
the case, and the detailed functional dependence of quantum coherence
on time almost always changes with $t$ in a complex manner, ruling out
any simple single-parameter characterization of spin decoherence.  To
be consistent with the standard literature, we often discuss or
describe our results by a single $T_2$, but we simply define $T_2$ as
the time it takes for the quantum memory to decay by a factor of $e$
(or the extrapolated time at which an approximate exponential decay form
will reach $1/e$).  This way we are not assuming any particular 
functional form of the quantum memory versus time decay.  
To be explicit, our results clearly indicate
the decay of the spin probability density over time.

The rest of the paper is organized as follows:
Section~\ref{SecSpecDiff} introduces the concept of spectral diffusion,
which is the only spin dephasing mechanism considered in this work (we
believe it to be the most important spin decoherence mechanism for
solid state quantum information processing using electron spins).
Sections~\ref{SecInteractions}, \ref{SecDeocherence}, and
\ref{SecDD} formally define the problem in terms of the
Hamiltonian, the decoherence measure, and pulse sequences,
respectively.
In Sec.~\ref{SecClusterExpansion},
we review our cluster expansion method\cite{witzelHahnShort,
  witzelHahnLong, witzelCDD} for solving this problem.
Section~\ref{shortTimeAndWavefunction} introduces
the role of confinement or wavefunction by considering the initial
time (``short time'') decay of spin coherence and then providing the
detailed theoretical considerations associated with the functional
form of the localized wavefunction as relevant for spin dephasing;
Section~\ref{SubSecContinuum} contains a
particularly important continuum approximation, which provides
convenient formulas that yield estimated $T_2$ times
as a function of the wavefunction size and shape.  
In Sec.~\ref{singleTriplet}, we consider
a specific recent experimental situation of singlet-triplet states in
a double quantum dot and show its equivalence to the single electron
case as far as spin dephasing is concerned; 
Sec.~\ref{Zamboni} discusses the limit on the experimentally
discovered Zamboni effect in enhancing spin coherence.
In Sec.~\ref{conclusion}, we conclude with a summary and a brief
discussion of open questions.

\section{Spectral diffusion}
\label{SecSpecDiff}

\begin{figure}
\begin{center}
\includegraphics[width=3.5in]{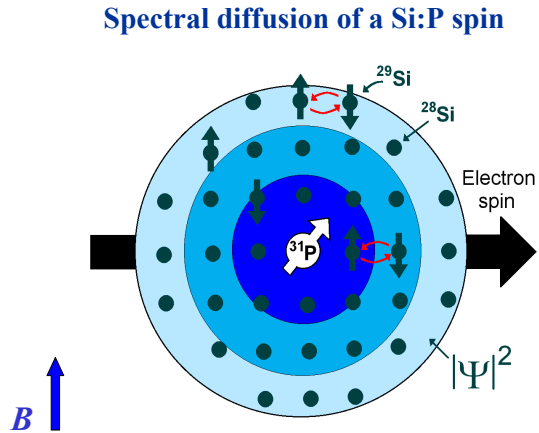}
\end{center}
\caption{
(Color online)
The electron of a P donor in Si experiences spectral
diffusion due to the spin dynamics of the enveloped bath of Si nuclei.
Of the naturally occurring isotopes of Si, only $^{29}$Si has a net
nuclear spin, which may contribute to spectral diffusion by
flip-flopping with nearby $^{29}$Si.
Natural Si contains about $5\%$ $^{29}$Si or less through
isotopic purification.  Isotopic purification or nuclear polarization
will suppress spectral diffusion in Si.
\label{SD_schematic}
}
\end{figure}

The spin decoherence
mechanism known as spectral diffusion (SD) has a long 
history\cite{herzog56, feher59, mims60_61, klauder62, zhidomirov69, chiba72},
and has been extensively
studied recently\cite{tyryshkin, abe, petta05, Delft,
desousa03b, witzelHahnShort, yaoHahn, witzelHahnLong, witzelCPMG, yaoCDD,
yaoCDDlong, witzelCDD, SaikinLinkedCluster} 
in the context of spin qubit
decoherence.
Consider a localized electron in a solid.  This is our central spin.
The electron spin could decohere
through a number of mechanisms.  In particular, spin relaxation would
occur via phonon or impurity scattering in the presence of spin-orbit
coupling, but these relaxation processes are strongly suppressed in
localized systems and can be arbitrarily reduced by lowering the
temperature and applying a strong external magnetic field,
creating a large electronic Zeeman splitting.  
In the dilute doping regime of interest in quantum
computation, where the localized electron spins are spatially well
separated, a
 direct magnetic dipolar interaction between the electrons
themselves is not an important dephasing mechanism.\cite{desousa03a}
Therefore, the 
interaction between the electron spin and the nuclear spin bath is
the important decoherence mechanism at low temperatures and
for localized electron spins.   Now we restrict ourselves to a
situation in the presence of an external magnetic field (which is the
situation of interest to us) and consider the spin
decoherence channels for the localized electron spin interacting with
the lattice nuclear spin bath.  Since the gyromagnetic ratios (and,
hence, the Zeeman energies) for the electron spin and the nuclear spins
are typically a factor of 2000 different (the electron Zeeman energy
being larger), hyperfine-induced direct spin-flip transitions
between electron and nuclear spins would be impossible 
(except as virtual transitions as will be discussed in Sec.~\ref{SecInteractions})
at low
temperature since phonons would be required for energy conservation.
This leaves the
indirect SD mechanism 
as the most 
effective electron
spin decoherence mechanism at low temperatures and finite magnetic
fields.  
The SD process is associated with the
dephasing of the electron spin resonance due to the temporally
fluctuating nuclear magnetic field at the localized electron site.
These temporal fluctuations cause the
electron spin resonance frequency to diffuse in the frequency space,
hence the name spectral diffusion.  
These fluctuations result from the dynamics of the nuclear spin bath
due to dipolar interactions with each other along with their
hyperfine interactions with the qubit.  This scenario is illustrated
in Fig.~\ref{SD_schematic}.
Nuclear spins in the spin bath flip-flop due to their mutual dipolar
coupling (since the typical experimental temperature scale $\sim
100~\mbox{mK}$ is essentially an infinite temperature scale for the
nuclear spins with nano-Kelvin scale coupling), and this leads to a
temporally random magnetic-field fluctuation on the central spin,
i.e., the electron.


\subsection{Interactions}
\label{SecInteractions}

In the most general form, the SD Hamiltonian for the central spin
decoherence problem may be written (in $\hbar = 1$ unit)
\begin{eqnarray}
\label{exactH}
\hat{\cal H} &=& \Omega_e \hat{S}_z + \sum_n \hat{\vec{I}}_n \cdot
\bm{A}_n \cdot \hat{\vec{S}} \\
\nonumber
&& {} + \sum_n \omega_n \hat{I}_{nz} + \sum_{n<m} \hat{\vec{I}}_n \cdot \bm{B}_{nm} \cdot \hat{\vec{I}}_m,
\end{eqnarray}
where $\hat{\vec{S}}$ and $\hat{\vec{I}}_n$ are vectors of spin
operators for the electron (central spin) and nucleus $n$
respectively, $\Omega_e$ and $\omega_n$ are their respective Zeeman
energies (with an external magnetic field applied in the $z$
direction), 
and $\bm{A}_n$ and $\bm{B}_{nm}$ are tensors describing
electronuclear and internuclear interactions respectively.
Isotropic Fermi-contact hyperfine (HF) interactions typically dominate
$\bm{A}_n$ (i.e., $\bm{A}_n = A_n \identity$) 
although anisotropic HF interactions,
due to dipolar contributions, may also be important.
Internuclear dipolar interactions often dominate $\bm{B}_{nm}$,
though other local interactions between nuclei such as
indirect exchange interactions\cite{shulman55, shulman58, sundfors69} may
also be significant.\cite{yaoHahn}
Typical energy scales are shown in Table~\ref{energyScales} for
convenience.

\begin{table}[h]
\caption[Interactions and estimated energy scales]{Interactions and estimated energy scales. (A similar table appears in Ref.~\onlinecite{yaoCDDlong}.)}
\begin{center}
\begin{tabular}{|c|c|c|c|}
\hline
Interaction & Symbol & Scale ($\hbar = 1$) & Scale ($k_B = 1$) \\
\hline \hline
Zeeman (electron) & $\Omega_e$ & $10^{11}~\mbox{s}^{-1}$~$^a$ &
$1~\mbox{K}$~$^a$ \\
\hline
Zeeman (nucleus) & $\omega_n$ & $10^{8}~\mbox{s}^{-1}$~$^a$ &
$1~\mbox{mK}$~$^a$
\\
\hline
Contact HF & $A_n$ & $10^{6}~\mbox{s}^{-1}$~$^b$ & $10~\mbox{$\mu$K}$~$^b$ \\
\hline
Cumulative HF & ${\cal A} = \sum_n A_n$ & $\lesssim 10^{11}~\mbox{s}^{-1}$~$^c$ &
$\lesssim 1~\mbox{K}$~$^c$ \\
\hline
Dipolar & $B_{nm}$ & $10^{2}~\mbox{s}^{-1}$ & $1~\mbox{nK}$ \\
\hline
Indirect exchange & $B_{nm}^{\mbox{\scriptsize Ex}}$ & $10^{2}~\mbox{s}^{-1}$ &
$1~\mbox{nK}$ \\
\hline
HF mediated & $A_{nm}$ & $10~\mbox{s}^{-1}$ & $10^{-1}~\mbox{nK}$ \\
\hline
\end{tabular}
\end{center}
\begin{flushleft}
$^a$ With a magnetic field of about $1~\mbox{T}$.\\
$^b$ In III-V compound quantum dots with $N \sim 10^5$ nuclei
  and also in Si:P donors. \\
$^c$ In III-V compound quantum dots, ${\cal A} \sim
  10^{11}~\mbox{s}^{-1}$.  In Si:P donors, ${\cal A} \sim
 f \times 10^{10}~\mbox{s}^{-1}$, where $f$ is the fraction of $^{29}$Si.
\label{energyScales}
\end{flushleft}
\end{table}

The HF energies are typically many orders of magnitude larger than
internuclear dipolar energies: $\|\bm{A}_n\| \gg \|\bm{B}_{nm}\|$.
By ignoring the $\bm{B}_{nm}$ term for a moment, 
decoherence may occur as a result of 
real or virtual electronuclear flip-flops
via the HF interaction.\cite{Basel, deng}
Such a process may be suppressed, however, by increasing the applied
magnetic field due to a conservation of Zeeman energies.  
These Zeeman energies are given by
$\Omega_e = \gamma_{S} B$ and
$\omega_n = -\gamma_{n} B$, where $B$ is the applied magnetic-field
strength, and $\gamma_{S}$ and $\gamma_{n}$ are
gyromagnetic ratios of the electron and nucleus $n$, respectively
(with $\gamma_{n}$ defined in an opposite sense as $\gamma_{S}$).
Typically, $\gamma_S \sim 10^3 \gamma_n$ so that the nuclear
Zeeman energy is negligible relative to the electron's Zeeman energy
and the electron must overcome its Zeeman energy barrier in order to flip.
In the limit of $\Omega_e \gg {\cal A}$, with ${\cal A} = \sum_n A_n$,
HF-induced electronuclear flip-flops are effectively suppressed.
When we reintroduce $\bm{B}_{nm}$, however, decoherence still occurs
and is well defined in the $\Omega_e \rightarrow \infty$ limit because
the electron spin will dephase as a result of nuclear field
fluctuations induced by inter-nuclear interactions.
This decoherence is spectral diffusion.

In the limit of a large applied field (formally we will say $\Omega_e
\rightarrow \infty$),
electron flips are completely suppressed.  
In this limit, the effective Hamiltonian becomes
$\hat{\cal H} \approx \sum_n \omega_n \hat{I}_{nz} 
+ \sum_n \left( \vec{A}_n \cdot \hat{\vec{I}}_n \right)  \hat{S}_z
+ \sum_{n<m} \hat{\vec{I}}_n \cdot \bm{B}_{nm} \cdot \hat{\vec{I}}_m$,
where $\vec{A}_n$ is the $z$ column vector of $\bm{A}_n$.
We are free to drop $\Omega_e
\hat{S}_z$ for any dynamical considerations now because it is a
conserved energy in this limit.
Effects due to anisotropic HF interactions may be treated
independently of SD with a trivial disregard to internuclear
interactions.\cite{witzelAHF}  For our purposes, therefore, we will
treat only the $z$ component of $\vec{A}_n$ from the Fermi-contact HF 
interaction, which leaves us with
\begin{eqnarray}
\label{HiFieldH}
\hat{\cal H} &\approx& 
\sum_n A_n \hat{I}_{nz}  \hat{S}_z
+ \sum_n \omega_n \hat{I}_{nz} + \sum_{n < m} \hat{\vec{I}}_n \cdot \bm{B}_{nm} \cdot \hat{\vec{I}}_m,~~~
\\
\label{contactHF}
A_{n} &=&
\frac{8\pi}{3}\gamma_{S}\gamma_{n}\hbar|\Psi(\bm{R}_{n})|^{2}.
\end{eqnarray}
Equation~(\ref{contactHF}) gives Fermi-contact HF coupling constants
that are proportional to the probability of the electron being at the 
nuclear site (the square of its wavefunction).

How large of a magnetic field must we apply for Eq.~(\ref{HiFieldH})
to be a valid effective Hamiltonian?
To be explicit, one may use quasidegenerate perturbation
theory~\cite{Winkler} to systematically transform Eq.~(\ref{exactH})
into the block form $\hat{\cal H} \approx \sum_{\pm} \lvert \pm
\rangle \hat{\cal H}_{\pm} \langle \pm \rvert$, where $\left\{ \lvert
+\rangle, \lvert-\rangle \right\}$ are the electron spin states.
This transformation will be convergent if $\Omega_e \gg {\cal A}$,
where ${\cal A} = \sum_n A_n$ is the maximum cumulative HF field.
In III-V compound quantum dots, $\Omega_e \sim {\cal A}$ with an
applied field of $1~\mbox{T}$, calling this approach into question; however,
the $\Omega_e \gg {\cal A}$ condition is overly strict
because this maximum HF field is never reached in a bath that is not
fully polarized.
It is more relevant to consider how higher orders of this
transformation [Eq.~(\ref{HiFieldH}) represents the zeroth order]
compare with the internuclear $\bm{B}_{nm}$ coupling term to
generate decoherence.
In the next order of the transformation,
HF-mediated interactions between nonlocal nuclei emerge.
This interaction, 
wellknown\cite{slichter, bloembergen55} as the
Ruderman-Kittel-Kasenya-Yosida interaction,
results from virtual electron spin flips and is suppressed with a
large applied magnetic field.
This next order contribution to the Hamiltonian is given by\cite{yaoHahn}
\begin{equation}
\label{HFM}
\HHFM = \sum_{n \ne m} A_{nm}
\hat{I}_{n+} \hat{I}_{m-} \hat{S}_z,
\end{equation}
where $A_{nm} = A_n A_m / \Omega_e$.
In applying this transformation, we must slightly rotate the basis
states; 
this results in a
``visibility'' loss\cite{shenvi05} of coherence estimated
as\cite{yaoHahn} $\sum_n \left(A_n / \Omega_e\right)^2$ and is
certainly small for a large bath in which
$\Omega_e \gtrsim {\cal A}$.
In this paper, we will restrict ourselves to the $\Omega_e
\rightarrow \infty$ limit and merely mention how HF-mediated or other
higher order interactions may play a role.

 Direct interactions between nuclei are
often dominated by their magnetic dipoles with the form 
\begin{equation}
\HD{n}{m} = \frac{\gamma_{n} \gamma_{m} \hbar}{2} 
\left[ \frac{\hat{I}_{n} \cdot \hat{I}_{m}}{R_{nm}^{3}} - \frac{3 (\hat{I}_{n} \cdot
    {\bm R_{nm}}) (\hat{I}_{m} \cdot {\bm R_{nm}})}{R_{nm}^5}\right].
\end{equation}
The dipolar interaction between nuclear spins in semiconductors 
has a typical strength of
$\HD{n}{m} \sim 10^2~s^{-1}$, which is much smaller than typical nuclear
Zeeman energies of about $10^8~s^{-1}$ in an applied field of $1~T$.
  Therefore, energy conservation arguments
allow us to neglect any term that changes the total Zeeman energy 
of the nuclei.
This leaves the following Zeeman energy-conserving 
secular\cite{abragam62, slichter} part of the interaction:
\begin{eqnarray}
\label{DipolarSecular}
\sum_{n < m} \HD{n}{m} 
& \approx & \frac{1}{2} \sum_{n \ne m} 
B_{nm} \left[\delta_{\gamma_n, \gamma_m} \hat{I}_{n+} \hat{I}_{m-} 
- 2 \hat{I}_{nz} \hat{I}_{mz}\right],~~~\\
\label{bnm}
B_{nm}&=&-\frac{1}{2}\gamma_{n}\gamma_{m}\hbar\frac{1 - 3 \cos^{2}{\theta_{nm}}}{R^{3}_{nm}},
\end{eqnarray}
where $R_{nm}$ is the length of the vector joining nucleus $n$ and
nucleus $m$, 
and $\theta_{nm}$ is the angle of this vector relative to the $z$
magnetic-field direction.
The $\delta_{\gamma_n, \gamma_m}$ denotes that
the flip-flop interaction between nuclei with 
different gyromagnetic ratios should be excluded 
because of Zeeman energy
conservation in the same way that the nonsecular part of the dipolar 
interaction is suppressed.
This occurs, for example, in GaAs because the two
isotopes of Ga and the isotope of As that are present have significantly
different gyromagnetic ratios.
This secular interaction corresponds to a $\bm{B}_{nm}$ matrix for
Eq.~(\ref{HiFieldH}),
with $B_{nm}$,
$B_{nm}$, and $-2 B_{nm}$ along the diagonal, respectively, in the
$x$-$y$-$z$ 
spin basis for a pair of nuclei having the same gyromagnetic
ratios.
(The $\hat{I}_{nz} \hat{I}_{mz}$ term plays an insignificant role,
which is why we use the same $B_{nm}$ symbol, in different fonts, for
the scalar and tensor).

In addition to the dipolar interactions between nuclei, an
indirect nuclear-spin exchange,\cite{bloembergen55, anderson55,
 shulman55, shulman58, sundfors69} which is
 mediated by virtual electron-hole
pairs, may also have a significant quantitative impact on SD in III-V
materials.\cite{yaoHahn}  
This interaction takes the form
\begin{equation}
\label{pseudoExchange}
\HEx{n}{m} = B^{\mbox{\scriptsize Ex}}_{nm} \hat{I}_n \cdot \hat{I}_m.
\end{equation}
The corresponding $\bm{B}_{nm}$ is $\bm{B}_{nm}^{\mbox{\scriptsize
    Ex}} =  B^{\mbox{\scriptsize Ex}}_{nm} \identity$.
The leading contribution to this pseudoexchange interaction 
for nearest neighbors may be expressed as~\cite{anderson55, shulman55}
\begin{equation}
\label{bnmEx}
B_{nm}^{\mbox{\scriptsize Ex}} = \frac{\mu_0}{4 \pi} 
\frac{\gamma_n^{\mbox{\scriptsize Ex}}
  \gamma_m^{\mbox{\scriptsize Ex}}}{{\bm R}_{nm}^3} \frac{a_0}{{\bm R}_{nm}},
\end{equation}
where $\gamma^{\mbox{\scriptsize Ex}}_n$ is the effective gyromagnetic ratio determined
by a renormalization of the electron charge density.\cite{yaoHahn}
This interaction has been experimentally studied many years 
ago.\cite{shulman55, shulman58, sundfors69}
In GaAs quantum dots, these interactions can be comparable to the
direct dipolar interactions.  There may be
other local interactions between nuclei in the bath, such as the
indirect pseudodipolar interaction~\cite{bloembergen55} or
internuclear quadrapole interaction, but the dipolar and indirect
exchange interactions alone account for nuclear magnetic resonance
line shapes.\cite{yaoHahn, shulman55, shulman58, sundfors69}  
In any case, all such local interactions may easily be included in
our formalism.

To summarize and put everything in a convenient general form, 
we approximate our Hamiltonian as 
$\hat{\cal H} \approx \sum_{\pm} \lvert \pm \rangle \hat{\cal H}_{\pm}
\langle \pm \rvert$ where
\begin{eqnarray}
\label{Hpm}
\hat{\cal H}_{\pm} &=& \pm\hat{\cal H}_{eb} + \epsilon \hat{\cal
  H}_{bb}.
\end{eqnarray}
$\hat{\cal H}_{eb}$ is the electron dependent part that plays the role of
coupling the electron spin to the bath, 
and $\hat{\cal H}_{bb}$ includes secular
(preserving nuclear Zeeman energy) 
bath-bath, i.e., internuclear,
interactions 
such as the secular dipolar interaction [Eq.~(\ref{DipolarSecular})] and the
exchange interaction [Eq.~(\ref{pseudoExchange})].
Zeeman energies are omitted from this Hamiltonian because they are
preserved by all included interactions and, thus, not relevant to the dynamics.

In the $\Omega_e \rightarrow \infty$ limit,
\begin{equation}
\hat{\cal H}_{eb} \approx \frac{1}{2} \sum_n A_n \hat{I}_{nz},
\end{equation}
with a factor of $1/2$ from the magnitude of the electron spin, and
$A_n$ given by Eq.~(\ref{contactHF}) with a proportionality to the square of the
electron wavefunction at site $n$.
To validate this approximate form, one may consider the effects of higher
order interactions of the canonical transformation, such as
the HF-mediated interactions [Eq.~(\ref{HFM})] 
that contribute to $\hat{\cal H}_{eb}$ (due to its $\hat{S}_z$
dependence).
These higher order interactions introduce additional decoherence
mechanisms that will shorten coherence times at lower magnetic
fields.  The main consideration of this paper is the decoherence
that may not be removed by simply increasing the magnetic-field
strength and is applicable in the limit of a large applied field 
at the point where decoherence is insensitive to the strength of the 
applied field.

\subsection{Decoherence}
\label{SecDeocherence}

We characterize decoherence as the expectation value of the electron
spin over the evolution of the experiment.  Since we consider a
dephasing-only Hamiltonian of the form
$\hat{\cal H} \approx \sum_{\pm} \lvert \pm \rangle \hat{\cal H}_{\pm}
\langle \pm \rvert$, we need only deal with dephasing decoherence.
Dephasing decoherence involves only the transverse component of the 
electron spin.
For a given experiment, we define the up and down evolution operators,
$\hat{U}^{\pm}$, as evolution operators for the bath given an
initially up or down electron spin.  If we just have free evolution
for a time $t$, then
$\hat{U}_{\pm} = \hat{U}^{\pm}_0 = \exp{\left(-i \hat{\cal H}_{\pm} t\right)}$.
In general, we can consider an experiment with a sequence of $\pi$
pulses that flip the electron spin between periods of free evolution
time $\tau_j$, so that
\begin{equation}
\hat{U}_{\pm} = ...
\exp{\left(-i \hat{\cal H}_{\mp} \tau_2\right)} \exp{\left(-i
  \hat{\cal H}_{\pm} \tau_1\right)}.
\end{equation}
The transverse component of the expectation value of the electron spin
will then decay, in magnitude, by a factor of 
$\left\| \left \langle \hat{U}_-^{\dag} \hat{U}_+ \right \rangle \right\| =
\left\| \langle \hat{W} \rangle \right\|$, where 
$\hat{W} \equiv \hat{U}_-^{\dag} \hat{U}_+$,
$\langle ... \rangle$ denotes an appropriately weighted
average over the bath states, and $\|...\|$ takes the magnitude of the
resulting complex number.
The coherence decay is thus characterized by 
$\left\| \langle \hat{W} \rangle \right\|$.  
In an echo experiment, one initializes an ensemble of 
pure electron spin state in some transverse direction, applies a
sequence of $\pi$ pulses designed to refocus the spins, and observes
an echo signal, $v_E = \left\| \langle \hat{W} \rangle \right\|$, at
the end of the experiment.

Given an arbitrary initial bath density matrix written in the form
$\hat{\rho}_{b} \equiv \sum_j P_j \lvert {\cal B}_j \rangle \langle
{\cal B}_j \rvert$, we may average $\hat{W}$ over bath states with
\begin{equation}
\langle \hat{W} \rangle = \sum_j P_j \langle {\cal B}_j \vert \hat{W}
\vert {\cal B}_j \rangle,
\end{equation}
where each $\lvert {\cal B}_j \rangle$ is a different bath state.
By referring to the energy scale estimates of Table~\ref{energyScales},
temperatures on the $\mbox{mK} - \mbox{K}$ scale are justifiably 
treated as infinite with respect to the $\mbox{nK}$ scale intrabath 
interactions.  The remaining Zeeman and Fermi-contact HF interactions in our
approximate Hamiltonian quantize the nuclear spins in the $z$
direction (involving only $\hat{I}_{nz}$ nuclear spin operators) and,
thus, we can approximate the initial bath density matrix as a mixed
state composed of uncorrelated pure nuclear spin states in this 
basis so that
\begin{equation}
\label{initialBathState}
\lvert {\cal B}_j \rangle \approx
\prod_{\otimes~n}
\left(\sum_{m} p_{nm} \lvert I_n, m \rangle_n \right),
\end{equation}
where $I_n$ is the magnitude of the $n$th nuclear spin and $\lvert
I_n, m \rangle_n$ represents the state of the $n$th nuclear spin with a
$z$ projection of $m$.
The cluster approximation in Sec.~\ref{clusterApprox} 
will make use of the assumption
that the bath is initially uncorrelated (at least, approximately).
Section \ref{shortTimeAndWavefunction} will use the $z$ quantization
as a further convenience.

\subsection{Dynamical decoupling pulse sequences}
\label{SecDD}

If we simply let the system freely evolve, the decay of the electron
spin expectation value strongly depends on the type of averaging we
perform over bath states.  If we consider an ensemble of electron
spins, each with its own bath, then we will see rapid electron spin
dephasing simply due to the distribution of HF nuclear fields, $\sum_n
A_n \hat{I}_{nz}/2$.  This is known as inhomogeneous broadening
because it broadens the electron spin precessional frequency due to
inhomogeneity of the effective magnetic field.
This, however, is an artifact of the ensemble or our lack of knowledge
of the effective nuclear field of a static bath and not true
decoherence.
If we consider a single electron spin with a known nuclear field,
or a homogeneous ensemble which may be obtained through
mode locking,\cite{modeLock} for example, we arrive at a dynamical
decoherence known as FID.  In our dephasing model,
FID can only arise from 
interactions among bath elements, such as local
dipolar or nonlocal HF-mediated interactions.

Traditionally,\cite{herzog56} nonhomogeneously broadened coherence is
measured from Hahn spin echoes.  The Hahn echo sequence simply
involves a single $\pi$ rotation midway through the evolution such that
$\hat{U}_{\pm}(t) = \hat{U}^{\pm}_1(t) =  \hat{U}^{\mp}_0(\tau)
\hat{U}^{\pm}_0(\tau)$, with $\tau = t/2$.
We denote this sequence with $\tau \rightarrow \pi \rightarrow \tau$:
free evolution for a time $\tau$, then a $\pi$ pulse, then free
evolution for a time $\tau$ again, with the arrows indicating the
sequence ordered in time.
This sequence will 
reverse the effect of any inhomogeneous static field.  
What remains
 is SD induced by a dynamical nuclear bath.
It is important to note, however, that the effects of the Hahn echo go
beyond the elimination of inhomogeneous broadening.  The Hahn echo is
also a DD sequence,\cite{longKhodjasteh} in 
which the first order of the Magnus expansion\cite{Magnus} is removed by the
fact that the time-averaged Hamiltonian, proportional to 
$\hat{\cal H}_+ + \hat{\cal H}_- \propto \hat{\cal H}_b$, decouples the qubit from the bath.
For this reason, the Hahn echo does not have the same effect as
homogeneous (or single qubit) free induction decay.\cite{yaoHahn} 
In particular, the Hahn echo removes the lowest-order effects of
HF-mediated interactions.\cite{yaoHahn}  
This is because HF-mediated interactions, having an $\hat{S}_z$ factor
[Eq.~(\ref{HFM})], belong to $\hat{\cal H}_{eb}$, and if we consider
only HF-mediated intrabath interactions, then $\hat{U}_1^{\pm} =
\exp{\left(\pm i \hat{\cal H}_{eb} \tau\right)} \exp{\left(\mp i
  \hat{\cal H}_{eb} \tau
  \right)} = \hat{\identity}$.

Given a DD sequence, such as the Hahn echo for a dephasing
system, coherence over a given net amount of time may be increased
through a rapid repetition of the basic sequence.  This strategy, known
as bang-bang in the quantum information community,\cite{ViolaPRL05}
gives coherence enhancement at the cost of more frequent applications
of $\pi$ pulses.  Pulses must be applied more frequently because
errors due to higher order terms of the Magnus expansion pile up over
the course of the sequence.  A better strategy is to use recursion,
rather than repetition, to generate concatenated
sequences.\cite{KhodjastehPRL}  
Such CDD, with the Hahn echo as the base sequence,
has been shown\cite{witzelCDD} to be
effective for the SD problem.  In fact, with each concatenation, we
demonstrated, in Ref.~\onlinecite{witzelCDD}, coherence enhancement with
an {\it increase} in the time between pulses (let alone, the net
sequence time).

With $l>0$ levels of concatenation, our CDD pulse sequence 
is recursively defined by \cite{KhodjastehPRL}
$\mbox{p}_l := 
\mbox{p}_{l-1} \rightarrow \pi \rightarrow \mbox{p}_{l-1} \rightarrow \pi$,
with $\mbox{p}_0 := \tau$.
CDD with $l=0$ is simply free evolution.  At $l=1$, we have 
$\mbox{p}_1 := \tau \rightarrow \pi \rightarrow \tau \rightarrow \pi$,
which is simply the Hahn echo (with an extra $\pi$ pulse at the end to
bring the electron spin back to its original phase apart from the
decoherence). 
With each concatenation, we do to the previous sequence what the Hahn
echo does to free evolution and, in this way, we obtain improved DD.
This sequence may be simplified by noting that two $\pi$ pulses in
sequence do nothing.  Thus,
\begin{equation}
\label{concatenatedPulseSequence}
\mbox{p}_l := \left\{ 
\begin{array}{ll}
\mbox{p}_{l-1} \rightarrow \pi \rightarrow \mbox{p}_{l-1} &,~ \mbox{odd}~l \\
\mbox{p}_{l-1} \rightarrow \mbox{p}_{l-1} &,~ \mbox{even}~l 
\end{array}\right.
\end{equation}
(again, arrows indicate sequences ordered in time)
and the up and down evolution operators at level $l$ have the recursive
form of\cite{yaoCDD}
\begin{equation}
\label{Ul_recursive}
\hat{U}_l^{\pm} = \hat{U}_{l-1}^{\mp} \hat{U}_{l-1}^{\pm}.
\end{equation}

Recently, a series of DD sequences was discovered by Uhrig\cite{UhrigSequence}
to be optimal in the number of pulses for the spin-boson model.
The $n$-pulse sequence in this series may be defined by
\begin{equation}
\label{UhrigTau}
\tau_j = \frac{1}{2} \left[\cos{\left(\frac{\pi (j-1)}{n+1}\right)} -
  \cos{\left(\frac{\pi j}{n+1}\right)}  \right]
\end{equation}
for $1 \leq j \leq n+1$.
Unlike CDD, Uhrig DD (UDD) requires only a linear (rather than
exponential) overhead in the number of pulses for each order of coherence enhancement.
Furthermore, UDD was shown\cite{UDD} to kill off successive orders in a time
expansion in a completely model-independent manner.
The UDD sequence has a strong advantage over CDD in its linear
versus exponential scaling of the number of pulses; however, it is
effective only when a time expansion is convergent, while
CDD is also effective in the intrabath perturbation for SD (this is
important since $B_{nm} \ll A_n$).

In Sec.~\ref{shortTimeAndWavefunction}, we will discuss the
wavefunction dependence in the short-time approximation.
It would be natural to discuss this in the context of the UDD series
since UDD is effective in this short-time limit.
However, the methods of Sec.~\ref{SubSecContinuum} work for the CDD
series but, unfortunately, do not work for the UDD series.
In this work, we therefore focus attention on FID and the CDD series.
Note, however, that the UDD and CDD series are the same for levels
zero (FID), 1 (Hahn\cite{herzog56}), and 2 (CPMG\cite{Meiboom, witzelCPMG}).

\section{Cluster method}
\label{SecClusterExpansion}

In this section, we review our cluster expansion
method\cite{witzelHahnShort, witzelHahnLong} for solving the SD
problem in a particularly simple and illuminating form.
Section~\ref{clusterApprox} gives the basic cluster expansion result,
the cluster approximation, in which we equate $\langle \hat{W}
\rangle$ [see Sec.~\ref{SecDeocherence}] to the exponentiation of
single-cluster contributions. 
 This is useful because perturbation
theory cannot be directly applied to $\langle \hat{W} \rangle$, but it
can be applied to its single-cluster contributions (it is the multicluster contributions
that are particularly problematic for perturbation theories due to the large number
of possibilities in combing different clusters).  We consider
aspects of such perturbation expansions in Sec.~\ref{SubSecSDPert}
that will be relevant to Sec.~\ref{shortTimeAndWavefunction}.

\subsection{Cluster approximation}
\label{clusterApprox}

The cluster expansion is based on the fact that our $\hat{\cal H}_{\pm}$
Hamiltonians
couple nuclei via relatively weak pairwise interactions in a large
bath with $N$ nuclei (any $n$-way interactions could justify such an
approximation as long as $n \ll N$).  The $\hat{W}$ operator, which is
the product of evolution operators arising from $\hat{\cal H}_{\pm}$, can,
in principle, be expanded into a sum of products of the Hamiltonian
interaction elements.  The interaction elements of each such term will
uniquely determine a set of clusters of nuclei;
each cluster, with respect to this term, may involve interactions
among itself but not among any other cluster, and no cluster may be
divided into further subclusters.  For example, a term with
$B_{1, 2} I_{1+} I_{2-} B_{3, 4} I_{3+} I_{4-}$ forms two clusters,
$\{1, 2\}$ and $\{3, 4\}$, while $B_{1, 2} I_{1+} I_{2-} B_{2, 3}
I_{2+} I_{3-}$ forms a single cluster of $\{1, 2, 3\}$.
The term ``cluster'' implies proximity among the member nuclei as
applicable to local dipolar interactions;
however, we may also treat nonlocal HF-mediated interactions
using the term cluster in a more general sense as a set of nuclei
that are interconnected by interactions under consideration.

If one considers a perturbative expansion of $\langle \hat{W} \rangle$
with respect to the pairwise interactions, one immediately faces the
problem that the number of terms of $\hat{W}$ scales in powers of $N$
with successive inclusion of the pairwise interactions, destroying any hope
of convergence when $N$ is large.
To resolve this problem, let us first partition $\hat{W}$ according to
the number of clusters involved in each term such that
\begin{equation}
\hat{W} = \sum_{p=0}^{N} \clusterCountW{p},
\end{equation}
where we define $\clusterCountW{p}$ as the sum of
the terms from $\hat{W}$ that involve $p$
independent clusters.  Note that $\clusterCountW{0} = \hat{\identity}$.
By considering only local interactions (e.g., dipolar), then it is
apparent that a perturbative expansion of $\langle \clusterCountW{1} \rangle$
with respect to the pairwise interactions does not suffer from the
adverse $N$ scaling suffered by $\langle \hat{W} \rangle$ because,
when interactions are local, there are $\bigO{N}$ clusters of any
size.  Even with nonlocal interactions, 
the perturbative expansion of $\langle \clusterCountW{1} \rangle$ is a
generally significantly better controlled expansion for large $N$ than that of
$\langle \hat{W} \rangle$.

Assume that, arising from a perturbative expansion with respect to the
pairwise interactions, $\clusterCountW{1}$ is well approximated 
when only including contributions due to clusters of some maximum size
that is much less than $N$.  
Along with our assumption that the bath nuclei are initially
uncorrelated, Appendix \ref{clusterFactorability} shows that
$\langle \clusterCountW{k} \rangle \approx 
\langle \clusterCountW{1} \rangle^k  / k!$
[Eq.~(\ref{clusterCountApprox})].  
This approximation breaks down as $k$ becomes significant relative to
$N$; however, for a large enough bath where the previous assumptions
are met,
\begin{equation}
\langle \hat{W} \rangle \approx \exp{\left(\langle \clusterCountW{1} \rangle\right)}
\end{equation}
and
\begin{equation}
\label{clusterApproxResult}
v_E = \left\| \langle \hat{W} \rangle \right\| \approx \exp{\left(\Real{\left\{\langle 
\clusterCountW{1} \rangle \right\}}\right)}.
\end{equation}
A formalized cluster expansion, with a discussion on convergence
tests, is presented in Ref.~\onlinecite{witzelHahnLong}.  The cluster
approximation presented in this subsection gives an equivalent result
with a simpler derivation.

\subsection{Single-cluster perturbation}
\label{SubSecSDPert}

Given the result of Eq.~(\ref{clusterApproxResult}), we have reduced
the problem of SD decoherence to that of perturbatively treating $\Real{\left\{\langle
  \clusterCountW{1} \rangle \}\right\}}$.  That is, we
wish to consider a perturbative expansion of $\hat{W}$ where we
neglect terms involving multiple clusters.  To be consistent with the
cluster approximation, such a perturbation should be directly or
indirectly tied to cluster size so that large clusters may be
neglected.  To this effect, we may perturbatively 
treat the pairwise interactions
(the intrabath perturbation) or we may consider a time
expansion which, in a sense, perturbatively 
treats all of the interactions (e.g.,
HF, dipolar, and HF mediated); in either case, larger clusters require more
interaction factors and thereby increase the order of the
perturbation.  

In this section, we will consider general perturbative properties that
apply to both the intrabath and time perturbation expansions.
 In Sec.~\ref{shortTimeAndWavefunction},
 we will specifically consider the time perturbation and see how it
 may be used in the formulation of a convenient continuum
 approximation.
In a perturbation whose order increases with increasing cluster size,
the lowest order of $\langle \hat{W} \rangle - 1$ is equivalent to the
same order of $\langle \clusterCountW{1} \rangle$ 
because terms of $\hat{W}$ with
multiple clusters are automatically higher order terms (products
of lower order terms).
To the lowest order, then, and in the context of   
CDD with $l$ levels of concatenation,
we will consider $\langle \hat{W} \rangle$.

By noting that 
$\hat{U}^{\pm}_l$ are unitary operators such that
$\left[\hat{U}^{\pm}_l\right]^{\dag} \hat{U}^{\pm}_l = \hat{\identity}$, we
may write
\begin{subequations}
\begin{eqnarray}
\Real{\left\{\left\langle \hat{W}_l \right\rangle\right\}} &=& 
\frac{1}{2} \left\langle \left[\hat{U}^{-}_l\right]^{\dag} \hat{U}_l^{+}
+ \left[\hat{U}_l^{+}\right]^{\dag} \hat{U}_l^{-} \right\rangle  \\
\label{DeltaDagDelta}
&=& 1 - \frac{1}{2} \left\langle\Delta_l^{\dag} \Delta_l \right\rangle,
\end{eqnarray}
\end{subequations}
where we define $\Delta_l \equiv \hat{U}^{+}_l - \hat{U}^{-}_l$.
Thus, $\left\langle\Delta_l^{\dag} \Delta_l \right\rangle$ gives a
measure of the decoherence.
By applying the recursive definitions for the $\hat{U}^{\pm}_l$ evolution
operators [Eq.~(\ref{Ul_recursive})],
\begin{equation}
\label{Delta_l_Def}
\hat{\Delta}_l \equiv \hat{U}_l^{+} - \hat{U}_l^{-} = \left[\hat{U}_{l-1}^{-},
  \hat{U}_{l-1}^{+} \right] = \left[\hat{U}_{l-1}^{-}, \hat{\Delta}_{l-1}\right],
\end{equation}
since $\hat{U}^{-}_{l-1}$ commutes with itself.

Let us consider a perturbation with a smallness parameter $\lambda$ in
which $\hat{U}_l^{\pm} = \hat{\identity} + \bigO{\lambda}$ for all $l \geq l_0$ for some $l_0$.
Two such perturbations are the time expansion with
 $\lambda = \tau$, $l_0 = 0$ 
(since no evolution occurs in the $\tau \rightarrow 0$ limit)
and intrabath perturbation
with $\lambda = \epsilon$,  $l_0 = 1$
(since there is a perfect spin echo refocusing in the $\epsilon
\rightarrow 0$ limit).
Because the identity commutes with anything,
it is easy to see from Eq.~(\ref{Delta_l_Def}) that
$\hat{\Delta}_l = \bigO{\lambda} \times \hat{\Delta}_{l-1}$ for all $l >
l_0$; this proves that we get successive cancellations of the low-order
perturbation ($\tau$ or $\epsilon$) 
with each concatenation of the sequence.\cite{witzelCDD}
The lowest order result is given by
\begin{equation}
\label{LowestOrderLevelRecursion_pre}
\hat{\Delta}_l
\approx
\lambda \left[
\left. \frac{d}{d \lambda} \hat{U}_{l-1}^{-}
\right\rvert_{\lambda = 0}
, \hat{\Delta}_{l-1}\right],~\forall~l>l_0.
\end{equation}
Conveniently, for all $l > l_0$,
\begin{eqnarray}
\label{SingleDerivativeRecursion}
\left. \frac{d}{d \lambda} \hat{U}_l^{\pm} \right\rvert_{\lambda = 0} &=&
\left. \frac{d}{d \lambda} \hat{U}_{l-1}^{+} \right\rvert_{\lambda = 0} +
\left. \frac{d}{d \lambda} \hat{U}_{l-1}^{-} \right\rvert_{\lambda = 0} \\
&=& 
\nonumber
2^{l-l_0} \left. \frac{d}{d \lambda} \left(\hat{U}_{l_0}^{+} + \hat{U}_{l_0}^{-}\right)/2
\right\rvert_{\lambda = 0},
\end{eqnarray}
so that Eq.~(\ref{LowestOrderLevelRecursion_pre}) becomes
\begin{equation}
\label{LowestOrderLevelRecursion}
\hat{\Delta}_l
\approx
\lambda
\left\{
\begin{array}{ll}
2^{l-l_0-1} \left[
\left. \frac{d}{d \lambda} \left(\frac{\hat{U}_{l_0}^{+} + \hat{U}_{l_0}^{-}}{2}\right)
\right\rvert_{\lambda = 0}
, \hat{\Delta}_{l-1}\right] &, ~l>l_0 \\
\left. \frac{d}{d \lambda} \left(\hat{U}_{l_0}^{+} - \hat{U}_{l_0}^{-}\right)
\right\rvert_{\lambda = 0}&,~l = l_0
\end{array}
\right..
\end{equation}
Note that in the $l=l_0+1$ case, Eq.~(\ref{Delta_l_Def}) yields
\begin{equation}
\hat{\Delta}_{l_0+1} \approx
\lambda^2
\left.
\left[
\frac{d}{d \lambda} \hat{U}_{l_0}^{-} 
,  \frac{d}{d \lambda} \hat{U}_{l_0}^{+}
\right]
\right\rvert_{\lambda = 0},
\end{equation}
which is equivalent to the corresponding case in
Eq.~(\ref{LowestOrderLevelRecursion}) recalling that
any operator commutes with itself.

\section{Wavefunction dependence in the short time limit}
\label{shortTimeAndWavefunction}

In this section, we use the formalism developed in
Sec.~\ref{SecClusterExpansion}, with the cluster approximation and general
perturbation formulation of Sec.~\ref{SubSecSDPert}, to derive results
applicable in a short-time limit and use these results to formulate a
continuum approximation useful for understanding the dependence of
spectral diffusion on the shape of the electron wavefunction.  
In Sec.~\ref{SubSecTimePert}, we apply the general results of
Sec.~\ref{SubSecSDPert} to obtain the lowest time perturbation
results of $\Real{\{\langle \clusterCountW{1} \rangle\}}$.  This is
done for the cases of free induction decay and concatenated echoes.
Section~\ref{pairCorrelations} shows how the nuclear dependent and
electron dependent parts of this lowest-order time
perturbation solution may be separated in a way that allows us to 
generically treat
the bath for any electron wavefunction.
Section~\ref{SubSecContinuum} takes this one step further by treating
the bath as a continuum so that we may obtain results via integration
for any given electron wavefunction.  
In Sec.~\ref{shortTimeJustification}, we discuss the
circumstances in which the time expansion may or may not be applicable.

\subsection{Time perturbation}
\label{SubSecTimePert}

Section~\ref{SubSecSDPert} considered the perturbations of concatenated
Hahn sequences in a general sense in the context of the cluster
approximation of Sec.~\ref{clusterApprox}.
Now we are specifically interested in the time expansion.  
We first address this in the case of FID
and then treat the Hahn sequence and its concatenations.

\subsubsection{Free induction decay}

The $l=0$ result of $\hat{\Delta} = \hat{U}_+ - \hat{U}_-$, without any
pulses, gives $\Delta_0 = -2 i \hat{\cal H}_{eb} \tau$.  Where
$\hat{\cal H}_{eb} = \sum_n A_n \hat{I}_{nz} / 2$, this result for
$\Delta_0$ is simply due to inhomogeneous broadening.  
In the case of free induction decay, 
we are not concerned with inhomogeneous broadening and would like
to obtain the SD decoherence of a single electron spin (or a
mode-locked ensemble\cite{modeLock}).  We may do this by entering the
rotating frame of reference for the electron precessing in the nuclear
field.  This may be done, effectively, by making the
transformations, $\hat{U}_0^{\pm}(t) \rightarrow \hat{U}_0^{\pm}(t) \exp{\left(\pm
  i \hat{\cal H}_{eb} t\right)}$.  for free induction decay, we then obtain
\begin{equation}
\label{lowestOrderTauFID}
\hat{\Delta}_0 = \left[\hat{\cal H}_{eb}, \hat{\cal H}_b \right] t^2.
\end{equation}
The free induction decay is then given by
Eq.~(\ref{clusterApproxResult}) via Eq.~(\ref{DeltaDagDelta}) with
this form of $\hat{\Delta}_0$.

\subsubsection{Concatenated dynamical decoupling}

We now consider the Hahn echo sequence and its concatenations: $l \geq
1$.  There is no need to go into a rotating frame as we did for FID
because these sequences automatically reverse the effects of
inhomogeneous broadening.
We will apply Eq.~(\ref{LowestOrderLevelRecursion}) by using $\lambda =
\tau$.
In the limit of $\tau \rightarrow 0$, no evolution can occur, so it is
apparent that $\left. \hat{U}_{0}^{\pm} \right\rvert_{\tau = 0} =
\hat{\identity}$ and, thus, $l_0 = 0$ in this context.
For $l = 1$, the Hahn echo [Eq.~(\ref{LowestOrderLevelRecursion})] becomes
\begin{eqnarray}
\label{lowestOrderTauHahn}
\hat{\Delta}_{1} &\approx&
 2 \left[\hat{\cal H}_{eb},
  \hat{\cal H}_{b} \right] \tau^2.~~~
\end{eqnarray}
Note that there is a simple relationship between
Eqs.~(\ref{lowestOrderTauHahn}) and (\ref{lowestOrderTauFID}) for the Hahn echo and
FID, respectively.  The relationship is not so simple when we move away
from the $\Omega_e \rightarrow \infty$ limit and consider HF-mediated
interactions.  In that situation, discussed in
Ref.~\onlinecite{yaoHahn}, 
the lowest order contribution to FID
in the time expansion will come from HF-mediated interactions but this
lowest order effect will be cancelled in the Hahn echo.

To consider concatenations of the echo, we simply 
apply the recursion of Eq.~(\ref{LowestOrderLevelRecursion}) to obtain
\begin{eqnarray}
\nonumber
\hat{\Delta}_l &=& -2^{(l^2 - l + 2) / 2}
\left[... \left[\left[\hat{\cal H}_{eb}, \hat{\cal H}_{b}\right], 
\hat{\cal H}_{b}\right], ...\right] \left(i \tau\right)^{l+1}\\
\label{lowestOrderTau}
&& {} + \bigO{\tau^{l+2}},~~\forall~l >
0,
\end{eqnarray}
with $l$ nested commutations abbreviated by $...$'s.
As a result of these nested commutations and
as observed in Ref.~\onlinecite{witzelCDD}, each concatenation introduces
larger cluster sizes to the lowest-order expression (i.e., each time
we commute with $\hat{\cal H}_{b}$, we may introduce an additional
nuclear site to any term of this operator).

\subsection{Pair ``correlations''}
\label{pairCorrelations}

A reasonable assumption for many solid-state spin baths is that the
bath Hamiltonian $\hat{\cal H}_{b}$, which excludes qubit-bath
interactions, is homogeneous.  That is, sites that are equivalent
in terms of the Bravais lattice are equivalent with regard to bath
interactions.  A notable exception to this is where isotopes in the
lattice are interchangeable; for example, three different isotopes of
Si may occupy any lattice site in Si, and two different isotopes of Ga
may occupy the Ga sublattice in GaAs.  However, if we simply want to know
the decoherence that results from averaging different types of
isotopic configurations, then we may regard the bath (apart from the
qubit interactions) as homogeneous and use isotopic probabilities in
expressions for $\hat{\cal H}_{b}$.  Then the only
inhomogeneity is in the interactions with the qubit, $\hat{\cal H}_{eb}$.
We can then factor out this inhomogeneous part and compute the rest in
a way that is independent of the qubit interactions.  This will be
convenient, for example, when analyzing a quantum dot in which the
wavefunction of the electron (whose spin represents the qubit) can
take on many shapes and sizes.

By referring to Eqs.~(\ref{lowestOrderTauFID}) and (\ref{lowestOrderTau}),
we can make the following factorization of the
homogeneous and nonhomogeneous parts of 
$\left\langle \hat{\Delta}_l^{\dag} \hat{\Delta}_l \right\rangle$
[determining SD via Eqs.~(\ref{DeltaDagDelta}) and (\ref{clusterApproxResult})]:
\begin{eqnarray}
\label{factoredOutElectron}
\frac{1}{2} \left. \left\langle\hat{\Delta}_l^{\dag} \hat{\Delta}_l\right\rangle
\right\rvert_{\Omega_e \rightarrow \infty} &=& 
\sum_{n, m}
A_n^{*} A_m 
f_{n, m}^{(l)} \tau^{p} \\
\nonumber
&& {} + \bigO{\tau^{p+2}},~~~~ 
\end{eqnarray}
where
\begin{equation}
\label{exponentVsConcatenations}
p(l) =
\left\{
\begin{array}{ll}
4 &,~l = 0 \\
2l+2 &,~l > 0
\end{array}\right..
\end{equation}
For $l = 0$ (FID), 
\begin{eqnarray}
\label{fnm0}
f_{n, m}^{(0)} &=& 
\frac{1}{2} \left\langle\left[\hat{F}_{nz}^{(0)}\right]^{\dag} \hat{F}_{mz}^{(0)}
\right\rangle,
\\
\hat{F}_{nz}^{(0)} &=& \left[ \hat{I}_{nz}, {\hat H}_{b} \right],
\end{eqnarray}
and for $l > 0$,
\begin{eqnarray}
\label{fnm}
f_{n, m}^{(l)} &\equiv&
(-)^{(l+1)}
2^{(l^2 - l + 1)}
\left\langle \left[\hat{F}_{nz}^{(l)}\right]^{\dag} \hat{F}_{mz}^{(l)} \right\rangle,~~~~ \\
\hat{F}_{nz}^{(l)} &=& \left[... \left[\left[\hat{I}_{nz}, {\hat H}_{b} \right], 
{\hat H}_{b}\right], ...\right].
\end{eqnarray}
where the $...$'s again denote $l$ nested commutations.

Since we assume the high field limit where secular coupling $\hat{\cal
  H}_b$ preserves nuclear polarization, 
$\left[\hat{\cal H}_b, \sum_n \hat{I}_{nz}\right] = 0$
so that $\sum_{m} f_{n, m}^{(l)} = \sum_{m} f_{m, n}^{(l)} = 0$ for
  any $n$.
Then, $f_{n, n}^{(l)} = -\sum_{m \ne n} \left( f_{n, m}^{(l)} + f_{m,
  n}^{(l)} \right)/2$. 
By using this fact, we may rewrite Eq.~(\ref{factoredOutElectron}) in
  terms of the differences of the HF constants; after all, if
  $A_n$ is the same for all nuclei, there is no nuclear induced 
spectral diffusion in the high field limit (nuclear flip-flops would have no
  effect on the electron).  We will assume that the HF constants are
  real, $A_n = A_n^*$, as is the case for the Fermi-contact interaction 
[Eq.~(\ref{contactHF})].  Then,
\begin{eqnarray}
\sum_{n, m} A_n A_m f_{n, m}^{(l)} &=& \frac{1}{2} \sum_{n \neq m} A_n A_m 
\left(f_{n,  m}^{(l)} + f_{m,  n}^{(l)}\right) \\
\nonumber
&& {} + \sum_n A_n^2 f_{n, n}^{(l)} \\
\nonumber
&=& -\frac{1}{4} \sum_{n \neq m} (A_n - A_m)^2 \left( f_{n, m}^{(l)} +
f_{m, n}^{(l)} \right).
\end{eqnarray}
Thus, in the short-time limit of Eq.~(\ref{factoredOutElectron}),
\begin{equation}
\label{factoredOutHFdiff}
\frac{1}{2} \left\langle\hat{\Delta}_l^{\dag} \hat{\Delta}_l\right\rangle
\approx - \frac{1}{2} \sum_{n \neq m} (A_n - A_m)^2 
\Real{\left\{f_{n, m}^{(l)}\right\}} \tau^p.
\end{equation}

The homogeneous part is represented by 
$f_{n, m}^{(l)}$, 
and by exploiting this
homogeneity, we note that this function is equivalent when we shift by
any Bravais lattice vector, $\Rvec$:
$f^{(l)}(\rVec_n, \rVec_m) \equiv f_{n, m}^{(l)} = 
f^{(l)}(\rVec_n - \Rvec, \rVec_m - \Rvec)$.
We may then relate any $\rVec_n - \Rvec$ to one of the basis sites of
the Bravais lattice, so then, with $b$ representing the corresponding
basis site of $\rVec_n$,
we may write
\begin{equation}
f_{n, m}^{(l)} = f_b^{l}(\rVec_n - \rVec_m).
\end{equation}

\subsection{Continuum approximation}
\label{SubSecContinuum}

The pair correlation formulation above is particularly convenient
in the context of a continuum approximation for HF coupling
constants.  From the Fermi-contact HF interaction
[Eq.~(\ref{contactHF})], 
\begin{equation}
A_n = \frac{8 \pi}{3} \gamma_e \gamma_n \hbar \left(\frac{d_n V_u}{a^3}\right) P(\rVec_n), 
\end{equation}
where $d_n$ is the charge density for the isotope at site $n$,
$V_u$ is the volume of the unit cell of the Bravais lattice, $a$ is
the lattice constant, and
$P(\rVec) \propto \|\Psi(\rVec)\|^2$ is the electron's probability
density normalized such that $\int d^3\rVec P(\rVec)/a^3 = 1$. 

Let $\ell$ characterize the correlation length scale from
Eq.~(\ref{fnm}).
Then, if $\left\lvert \partial_i P \right\rvert
/ \left\lvert \partial^2_{ij} P \right\rvert \ll \ell$, where
$\partial_i = \partial / \partial x_i$ and 
$\partial^2_{ij} = \left(\partial / \partial x_i\right) \left(\partial
/ \partial x_j\right)$, for a given $n$ and $m$ pair in 
Eq.~\ref{factoredOutElectron} we may use
\begin{equation}
P(\rVec_n) - P(\rVec_m) \approx
\left(\rVec_n - \rVec_m\right) \cdot \vec{\nabla} 
P\left(\rVec_m\right).
\end{equation}
Furthermore, by using a continuum approximation where we replace one of the summations
with an integral, Eq.~(\ref{factoredOutHFdiff}) then becomes
\begin{equation}
\label{continuum1}
\frac{1}{2} \left\langle\hat{\Delta}_l^{\dag} \hat{\Delta}_l\right\rangle
\approx  - \frac{\tau^p}{2} \sum_{j} C_{j}^{l} \int \frac{d^3 \rVec}{a^3} 
\left[\rVec_j \cdot \vec{\nabla} P\left(\rVec\right)\right]^2,
\end{equation}
where
\begin{equation}
\label{Cjl}
C_{j}^{(l)} = n_c
\left\langle
\left[\frac{8 \pi}{3} \gamma_e 
\gamma_I \hbar d_I \left( \frac{V_u}{a^3} \right) \right]^2 f_{I}^{(l)}(\rVec_j)\right\rangle_{I}
\end{equation}
and $n_c$ is the number of sites in the conventional cell (of volume
 $a^3$) and $\langle ... \rangle_{I}$ averages over different
 isotopes.
Note that $n_c$ is equal to the number of basis sites multiplied by
 $a^3 / V_u$.

By rearranging Eq.~(\ref{continuum1}),
\begin{eqnarray}
\nonumber
\frac{1}{2} \left\langle\hat{\Delta}_l^{\dag} \hat{\Delta}_l\right\rangle
&\approx& \tau^p
 \int \frac{d^3 \rVec}{a^3} 
\left[\vec{\nabla} P(\rVec)\right]\transpose
{\bf M}^{(l)}
 \left[\vec{\nabla} P(\rVec)\right], \\
\label{continuum2}
{\bf M}^{(l)} &=& - \frac{1}{2} \sum_{j} \left[\rVec_j\right]
\left[\rVec_j\right]\transpose C_j^{(l)},
\end{eqnarray}
where $\vec{x}\transpose$ denotes the transpose of any vector $\vec{x}$,
$\left[\rVec_j\right] \left[\rVec_j\right]\transpose$ is an outer
product,
 and ${\bm M}^{(l)}$ is a
matrix.  
It is important to note that ${\bf M}^{(l)}$ is independent of
the electron wavefunction (or its probability density); these are
constants that are predetermined for a particular lattice and applied
magnetic-field direction.  The
wavefunction dependence is entirely of the form
$\int d^3 \rVec \left[\partial_i P(\rVec)\right] \left[\partial_j
  P(\rVec)\right]$.
Being symmetric, we may diagonalize ${\bf M}^{(l)}$ to the form
${\bf M}^{(l)} = \sum_i \vec{u}_i \vec{u}_i\transpose m_i^{(l)}$ so that
\begin{equation}
\frac{1}{2} \left\langle\hat{\Delta}_l^{\dag} \hat{\Delta}_l\right\rangle
\approx
\tau^p \sum_i m_i^{(l)} \int \frac{d^3 \rVec}{a^3} 
 \left[\vec{u}_i \cdot \vec{\nabla} P(\rVec)\right]^2.
\end{equation}
Details of how we actually computed ${\bm M}^{(l)}$ for various
systems are given in Appendix~\ref{computingM}.

Putting this in yet another form,
\begin{eqnarray}
\label{shortTimeBehavior}
\frac{1}{2} \left\langle\Delta_l^{\dag} \Delta_l \right\rangle &=& \left[\tau /
\tau_0^{(l)}\right]^{p(l)} + \bigO{\tau^{p(l)+2}}, \\
\label{tau0}
\frac{1}{\left[\tau_0^{(l)}\right]^{p(l)}} &\approx& \sum_i 
\frac{\int d^3\rVec 
\left[\vec{u}_i \cdot \vec{\nabla} P(\rVec)\right]^2 / a}
{\left[\mu_i^{(l)}\right]^{p(l)}},~~~~ 
\end{eqnarray}
where $p(l)$ is defined by Eq.~(\ref{exponentVsConcatenations}) and
the $\mu_i^{(l)} = 
\left[m_i^{(l)} / a^2 \right]^{-1/(2l+2)} $ have
units of time.  
For a given concatenation level $l$, the echo signal
[Eq.~(\ref{clusterApproxResult})] is approximately
\begin{equation}
\label{initEchoDecay}
v_E \approx \exp{\left(-\left[\tau/\tau_0^{(l)}\right]^{p(l)}\right)}
\end{equation}
in the limits of a strong applied
magnetic field and in the short-time approximation (i.e., extrapolated
from the short-time behavior which may or may not be valid at $\tau =\tau_0$).

We will now treat, specifically, the case of a quantum well with
thickness $z_0$ and
Fock-Darwin radius $r_0$ (resulting from a combination of parabolic
confinement and confinement due to the magnetic field).\cite{desousa03b}
In this case, the wavefunction is sinusoidal in the $z$ direction and
has a Gaussian form in the lateral direction.  The probability density
is of the form
\begin{eqnarray}
\nonumber
P(x, y, z) &\propto& \exp{\left(- \frac{x^{2} +
    y^{2}}{r_0^2}\right)} \times \\
\label{gaussianDotProb}
&& \cos^{2}{\left(\frac{\pi}{z_0} z \right)} \Theta(z_0/2 - |z|).
\end{eqnarray}
Let us consider the case 
where the $z$ vector is an eigenvector of $\bm{M}^{(l)}$
(e.g., when the problem, with the applied magnetic field direction,
is symmetric about the $z$ axis).
By recalling that $P(\rVec)$ should be normalized such that $\int d^3\rVec
P(\rVec)/a^3 = 1$, 
\begin{eqnarray}
\int d^3\rVec 
\left[\frac{\partial P(\rVec)}{\partial x} \right]^2
&=& \int d^3\rVec 
\left[\frac{\partial P(\rVec)}{\partial y} \right]^2 \\
\nonumber
&=& \frac{3 a^6}{4 \pi z_0 r_0^4}, \\
\int d^3\rVec 
\left[\frac{\partial P(\rVec)}{\partial z} \right]^2
&=& \frac{\pi a^6}{z_0^3 r_0^2}.
\end{eqnarray}
We may simply plug these into Eq.~(\ref{tau0}) as follows:
\begin{equation}
\label{tau0gaussian}
\tau_0^{(l)} =
\left\{\frac{\pi (a / z_0)^2}{N_a} \left[\mu_z^{(l)}\right]^{-p(l)}
+ \frac{3 (a / r_0)^2}{2 \pi N_a} \left[\mu_x^{(l)}\right]^{-p(l)}
\right\}^{\frac{-1}{p(l)}},
\end{equation}
defining $N_a = V / a^3$ as the number of $a \times a \times a$
lattice cubes in a quantum dot of volume $V = z_0 r_0^2$.
When $r_0 \gg z_0$, for example, we have
\begin{equation}
\label{tau0gaussianFlatDot}
\tau_0^{(l)} \approx \mu_z^{(l)} \left[
  \frac{N_a (z_0/a)^2}{\pi}\right]^{\frac{1}{p(l)}}.
\end{equation}

Table \ref{GaAsInAs_mu} shows computed values of $\mu_i^{(l)}$ 
for the GaAs or InAs lattice with $0 \leq l \leq 3$ (free induction decay,
Hahn echo, and up to three levels of concatenation). 
Subscripts of $\mu$ in this table indicate corresponding eigenvector, $\vec{u}_i$,
lattice directions.
GaAs and InAs both have a zinc-blende structure with the $^{75}$As
atoms on one of the two fcc lattices and respective lattice constants
of $5.65$ and $6.06$~\AA.\cite{kittel}  
The natural abundance of Ga isotopes $60.4\%$~$^{69}$Ga and 
$30.2\%$~$^{71}$Ga, and the natural abundance of In isotopes are 
$4.3\%$~$^{113}$In and $95.7\%$~$^{115}$In.\cite{chem_handbook}  
The gyromagnetic ratios are
$\gamma_{I} = (4.60, 8.18, 6.44, 5.90, 5.88) \times 10^3 \mbox{(s G)$^{-1}$}$ 
for
$^{75}$As, $^{71}$Ga, and $^{69}$Ga, $^{113}$In, and $^{115}$In respectively.\cite{gyroRef}
We have also used the following respective charge densities 
$d(I) = (9.8, 5.8, 5.8, 2.3, 2.3) \times 10^{25}~\mbox{cm}^{-3}$; these
charge densities were estimated in Ref.~\onlinecite{paget77} for GaAs
and estimated in Ref.~\onlinecite{yaoCDDlong} using the technique in Ref.~\onlinecite{paget77} for
InAs.
The Ga and As nuclei have spin magnitudes of $I=3/2$ and the In nuclei
have spin magnitudes of $I=9/2$, which we account
for properly.
The table shows the results for an applied magnetic field in the $[001]$
or $[110]$ lattice directions.
For the bath-only Hamiltonian $\hat{\cal H}_{B}$, we have included the
secular dipolar interaction [Eq.~(\ref{DipolarSecular})] and, for GaAs, indirect exchange
interaction [Eq.~(\ref{pseudoExchange})] with\cite{sundfors69}
$B_{nm}^{\scriptsize Ex} = 
-\gamma_{n}\gamma_{m} \hbar ( \sqrt{2.6}~\mbox{\AA} / 2 R_{nm}^4)$.
The table shows GaAs results when we include or exclude indirect
exchange interactions; the remainder of the table only considers
spectral diffusion induced by dipolar interactions among bath nuclei.
Because of the near degeneracy of the In gyromagnetic rations and the
natural predominant abundance of $^{115}$In, we simply show results
for $100\%$ $^{115}$In; this should also yield the lower bound of
dipolar-induced spectral diffusion decoherence times for any InAs/GaAs
mixture (since In induces the strongest decoherence due to its large
$I=9/2$ spin).  Mixing a little Ga into InAs or even mixing a little
In into GaAs will increase decoherence times
as a result of the reduced probability for any given cluster to be of
the same nuclear isotope (only these clusters can contribute in the
high magnetic field limit).  Contributions from the As nuclei to
decoherence is unavoidable in any such mixture; we thus give As-only
results in the table as an upper bound for decoherence times in
InAs/GaAs mixtures.  This As contribution, however, will vary depending upon
lattice constant; we give the full range in which slightly
longer decoherence times result from using the slightly larger InAs
lattice constant and vice-versa for the GaAs lattice constant.  

\begin{table*}[t]
\caption{
GaAs and/or InAs material $\mu_{i}$ parameters [see Eqs.~(\ref{tau0}) and
  (\ref{initEchoDecay})] in microseconds with the
  magnetic field, $B$, in the $[001]$ or $[110]$ lattice directions.
Considers only dipolar interactions except for values in parentheses
that also include indirect exchange.
}
\begin{center}
\begin{tabular*}{0.75\textwidth}{@{\extracolsep{\fill}}lccccc}
\hline
\hline
& \multicolumn{2}{c}{$B~||~[001]$} & \multicolumn{3}{c}{$B~||~[110]$}
\\
\cline{2-3}
\cline{4-6}
Level & $\mu_{[100], [010]}$ & $\mu_{[001]}$ 
& $\mu_{[110]}$ & $\mu_{[1\overline{1}0]}$ & $\mu_{[001]}$ \\
\hline
\multicolumn{6}{c}{GaAs with natural isotope abundances} \\
$l=0$ & 0.36~(0.29) & 0.33~(0.37) & 0.36~(0.28) & 0.28~(0.31) & 0.41~(0.30)\\
$l=1$ & 0.25~(0.21) & 0.23~(0.26) & 0.26~(0.22) & 0.20~(0.20) & 0.29~(0.22) \\
$l=2$ & 2.2~(1.8) & 2.0~(2.1) & 2.1~(1.7) & 1.7~(1.8) & 2.2~(1.8)\\
$l=3$ & 4.1~(3.6) & 3.6~(3.7) & 3.8~(3.2) & 3.2~(3.3) & 3.9~(3.2) \\
\hline
\multicolumn{6}{c}{InAs with $100\%$ $^{115}$In} \\
$l=0$ & 0.29 & 0.26 & 0.30 & 0.23 & 0.34\\
$l=1$ & 0.20 & 0.19 & 0.21 & 0.16 & 0.24 \\
$l=2$ & 1.6 & 1.4 & 1.5 & 1.2 & 1.6\\
$l=3$ & 2.7 & 2.4 & 2.6 & 2.2 & 2.6 \\
\hline
\multicolumn{6}{c}{As only (upper bound for GaAs/InAs mixtures)} \\
$l=0$ & 0.49--0.55 & 0.45--0.50 & 0.50--0.56 & 0.38--0.43 & 0.57--0.63\\
$l=1$ & 0.35--0.39 & 0.32--0.35 & 0.36--0.40 & 0.27--0.30 & 0.40--0.45 \\
$l=2$ & 3.3--3.8 & 2.9--3.4 & 3.2--3.6 & 2.6--2.9 & 3.2--3.7\\
$l=3$ & 6.5--7.5 & 5.6--6.5 & 6.2--7.1 & 5.1--6.0 & 5.8--7.1 \\
\hline
\hline
\end{tabular*}
\end{center}
\label{GaAsInAs_mu}
\end{table*}

It is important to note that the decay time for the overall CDD pulse
sequence at level $l$ is $t_0 = 2^l \tau_0$.  Values of $\tau$
represent the time between pulses rather than the overall time.  Thus,
the fact that the values of $\mu_i$ [related to $\tau_0$ via
  Eq.~(\ref{tau0})] increase in Table~\ref{GaAsInAs_mu} as $l$ increases
beyond $l=1$ yields extra coherence time enhancement beyond the $2^l$
extension of $t_0$.  That is, as noted in Ref.~\onlinecite{witzelCDD}, not
only does concatenation increase the net coherence time, but it also
{\it decreases} the frequency at which one must apply pulses in order
to maintain coherence.

\begin{figure}
\begin{center}
\includegraphics[width=3in]{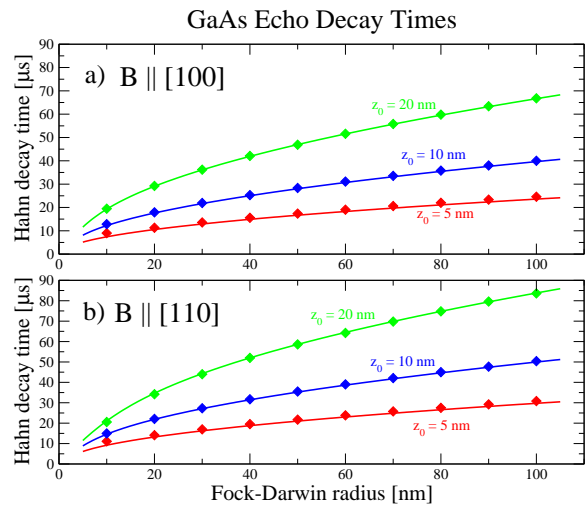}
\end{center}
\caption{
(Color online)
Hahn echo total decay times, $t_0^{1} = 2 \tau_0^{1}$, for GaAs quantum dots with various Fock-Darwin
radii $r_0$ and thickness $z_0$ comparing the continuum approximation
results (solid lines) to $1/e$ decay times obtained via the cluster
expansion and previously presented in
Ref.~\onlinecite{witzelHahnLong}.  This example includes dipolar but
not indirect exchange interactions; it therefore slightly
over-estimates the decay time.
\label{FigGaAsContinuum}}
\end{figure}

Figure \ref{FigGaAsContinuum} demonstrates the accuracy of the
continuum approximation compared with exact cluster calculations for
Gaussian shaped quantum dots.  It also shows that the continuum
approximation is best for large dots and deviates as we consider
smaller dots.  In fact, it has been clearly reasoned\cite{desousa03b} that the
decay time must approach infinity as the quantum dot size approaches
zero extent, but the continuum approximation fails to capture this trend.

\subsection{When is the short-time limit appropriate?}
\label{shortTimeJustification}

The short-time behavior will be exhibited on a time scale that is
short relative to the time scale of the dynamics of the relevant
cluster contributions in $\clusterCountW{1}$.  Because a large number
of these cluster contributions are added together in
$\Real{\left\{\langle \clusterCountW{1}\rangle \right\}}$
 [and then exponentiated to yield $v_{E}$ from
  Eq.~(\ref{clusterApproxResult})], it is possible for the decay of the
echo, $v_{E}$, to occur on a time scale that is small relative to the
dynamical time scales of any significantly contributing cluster.  In
particular, the decay will exhibit a short-time behavior when the
clusters with the fastest dynamics dominate 
$\Real{\left\{\langle \clusterCountW{1}\rangle \right\}}$.  When there
is a mixture of dynamical time scales playing a role, then the 
short-time behavior will be washed out by oscillations generated by
HF-induced precessions of the nuclei.

In considering the dynamical time scale of a cluster contribution, we
really want to know the effect of this cluster on electron spin
dephasing.  This is determined by the difference in HF energies
(with a reciprocal relationship between time and energy)
for different spin polarization configurations of the nuclei in the
cluster.  
In the extreme case that all of the
HF energies are the same, there is no spectral diffusion
induced (the dynamical time scale is infinite).  
The dynamical time scale is also determined by the
interactions between the nuclei that can cause changes in the spin
polarization configurations (turning these interactions off will also
shut off spectral diffusion); however, we can estimate a lower bound
time scale from just the inverse of differences in HF energies among
the cluster.

\begin{figure}
\begin{center}
\includegraphics[width=3in]{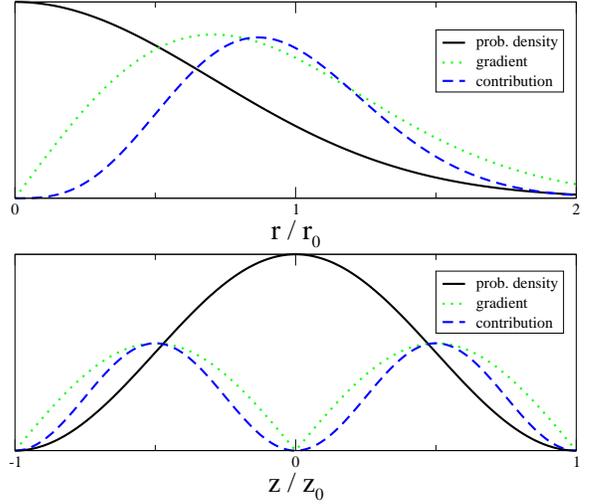}
\end{center}
\caption{
(Color online)
A comparison of the position dependence of the electron probability density,
absolute value of the gradient of this density [strictly speaking, 
$\|\frac{\partial}{\partial r} P(\vec{r}) \|$ and 
$\|\frac{\partial}{\partial z} P(\vec{r}) \|$ for the top and bottom
graphs respectively], and the resulting contribution to the short-time
behavior of SD
[integrand of Eq.~(\ref{tau0})
$\propto r \left(\frac{\partial}{\partial r} P(\vec{r})\right)^2$
and  $\propto \left(\frac{\partial}{\partial z} P(\vec{r})\right)^2$
for the top and bottom graphs respectively]
for a laterally confined quantum dot of the form of Eq.~(\ref{gaussianDotProb}).
The curves have arbitrary vertical scales.
The short-time behavior exhibited by the SD decay of such quantum dots
is related to the fact that the maximum contributions occur roughly
where
the gradient is maximum and differences in HF coupling among
neighboring nuclei are large.
\label{FigGaussianWF}}
\end{figure}

\begin{figure}
\begin{center}
\includegraphics[width=3in]{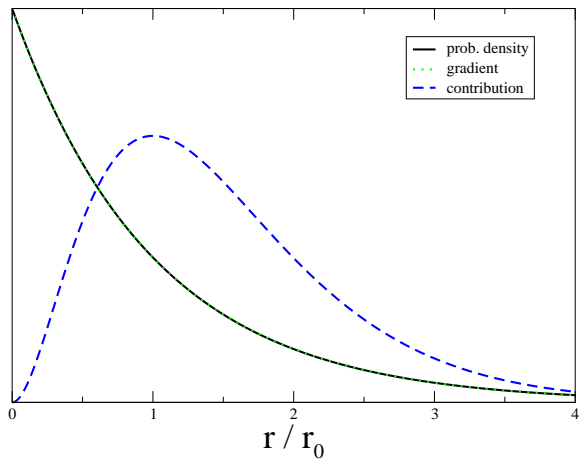}
\end{center}
\caption{
(Color online)
A comparison of the position dependence of the electron probability density,
absolute value of the gradient of this density, $\|
\frac{\partial}{\partial r} P(\vec{r}) \| \propto P(\vec{r})$,
and the resulting contribution to the short time behavior SD
[integrand of Eq.~(\ref{tau0})
$\propto r^2 \left(\frac{\partial}{\partial r} P(\vec{r})\right)^2$]
for an electron with an exponential probability density of the form
$P(\vec{r}) = \exp(-r/r_0)$ where $r = \|\vec{r}\|$.
Donor bound electrons, such as in Si:P, have an exponential-type decay
(though not quite as simple).
The curves have arbitrary vertical scales.
The failure of the time expansions in such systems is related to the fact
that the maximum contributions do {\it not} occur where the gradient
is maximum.
\label{FigExpWF}}
\end{figure}

We first address the $\Omega_e \rightarrow \infty$ limit and later
discuss, briefly, the short-time behavior of HF-mediated interactions.
Our previous\cite{witzelHahnShort, witzelHahnLong, witzelCPMG}
results in the $\Omega_e \rightarrow \infty$ limit 
show that the echo decay typically exhibits a short-time
behavior in quantum dots with assumed Gaussian shaped
wavefunctions but not in donors with exponential-like wavefunctions.
This is understood in the following way.  In the case of the donor,
the fastest dynamics come from those few nuclei in the center that
have large differences in HF energies.  These are too few in
number to dominate the decay; therefore, a mixture of time scales must
play a role, slowing down but contributing more as we consider
clusters further from the center, and the short-time behavior is
washed out.  For a quantum dot with a Gaussian shaped wavefunction,
however, the fastest contributions occur where the wavefunction
gradient is large in a ring around the dot at a radius of the
characteristic size of the dot.  There are many such clusters that are
collectively capable of dominating the echo decay so that it will
exhibit a short-time behavior.
These arguments are illustrated in Figs.~\ref{FigGaussianWF} and \ref{FigExpWF}.

There is a simple self-consistent check of the short-time behavior
[Eqs.~(\ref{shortTimeBehavior}) and (\ref{tau0})] in the 
$\Omega_e \rightarrow \infty$ limit by comparing $\tau_0$ to the
fastest dynamical timescale estimated by the maximum gradient
multiplied by the lattice spacing (as a typical distance scale between
nuclei).  
Thus, the short-time behavior is valid when $\tau_0 \ll 
1 / \left[a \max{\lvert \vec{\nabla} A(\vec{r}) \rvert}\right]$.  For
an electron probability density of the form of
Eq.~(\ref{gaussianDotProb}), $\max{\lvert \vec{\nabla} A(\vec{r})
  \rvert} \sim \max{\left(A_n / r_0, A_n / z_0 \right)}$.  With $r_0 \gg
z_0$, the dynamical time scale has more to do with the position of the
cluster in the $z$ direction rather than the radial direction with a
time scale estimate of $\tau \sim z_0 / a A_n$.  With a $10~\mbox{nm}$ quantum
dot thickness for $z_0$ and lattice constant of about $0.5~\mbox{nm}$, this
sets a time scale of about $10~\mbox{$\mu$s}$.  This estimation is
slightly pessimistic because we have computed cluster expansion
results for $z_0 = 10~\mbox{nm}$ and $r_0 = 100~\mbox{nm}$ to obtain
$\tau_0 \sim 25~\mbox{$\mu$s}$ Hahn echo decay exhibiting the 
short-time behavior even though $\tau_0 > 10~\mbox{$\mu$s}$.  However, our
$10~\mbox{$\mu$s}$ estimate is expected to be overly short because the
dynamics is slowed by the weak $10~\mbox{ms}$ dipolar coupling
(necessarily involved in any decoherence in the $\Omega_e \rightarrow
\infty$ limit).

This paper is primarily restricted to the $\Omega_e \rightarrow
\infty$ regime.  For the moment, however, let us consider a finite
$\Omega_e$ and discuss the question of short-time behavior validity
with HF-mediated interactions.
Because these are nonlocal interactions,
differences in HF energies can be as large as the HF
energies themselves.  This sets a dynamical time scale on the order of
$1/A_n \sim 1~\mbox{$\mu$s}$.  Also, the fact that HF-mediated
interactions are weak does not help too much in this case because
there are $N \gtrsim 10^5$ nuclei with which a given nucleus may
interact; this yields a strong collective interaction of about $N
A_{nm} \sim A_n$.  In a manner of thinking, on a microsecond time scale, a
given nucleus is likely to flip-flop with some other nucleus in the
bath through the HF-mediated interaction.  On the other hand,
flip-flops between nuclei with large differences in HF energies
will be suppressed due to HF energy conservation.  For this reason,
clusters with the fastest dynamics, those with large HF energy
differences, will give weak contributions due to energy conservation,
while the larger contributors with matching HF energies will
have slow dynamics.  This leads to a situation similar to that of the
donor electron in the $\Omega_e \rightarrow \infty$ limit, where the
fastest contributors cannot dominate and, therefore, the short-time
behavior is washed out.  A proper treatment of HF-mediated
interactions, therefore, would not use the time perturbation; instead,
we should use the intrabath perturbation, perturbatively
treating both dipolar and
HF-mediated interactions with respect to the HF
interaction.  

The short-time behavior that we consider in this paper, with its
convenient application in the continuum approximation, only applies in
the $\Omega_e \rightarrow \infty$ limit, where the HF-mediated
interactions are negligible.  How large does $\Omega_e$ need to be
for this limit to apply?  We may pose this question differently
to ask: How does $\Omega_e$ affect the time scale of short-time
behavior convergence?  If we can push this time scale sufficiently
larger than $\tau_0$, the short-time behavior will emerge.  
The HF-mediated interaction represents the lowest-order interaction
with a magnetic-field dependence, $A_{nm} = A_n A_m / 4 \Omega_e$.  
The time scale of the fast
dynamics due to the HF-mediated interactions will be a combination of
the difference in HF energies ($\sim A_n$) and the collective
HF-mediated energy ($\sim N A_{nm}$) to give a time scale of
$\sim \left(A_n N A_{nm}\right)^{-1/2} \propto \sqrt{\Omega_e}$.
Therefore, according to this simple argument,
it is necessary to quadruple $\Omega_e$ in order to double the
time scale of the short-time behavior as long as HF-mediated
interactions are dominant over local internuclear interactions.
The lowest-order effects of HF-mediated interactions, however, 
are reversed by any DD refocusing technique.
Therefore, this argument is only relevant for FID.

\section{Singlet-triplet double quantum dots}
\label{singleTriplet}

Remarkable experiments have recently\cite{petta05, Delft} investigated the
coherence properties of a {\it single} qubit in GaAs quantum dots.
In the earlier of these experiments,\cite{petta05}
 the qubit was not the spin of a
single electron but, rather, a subspace of two electron spins, each in
separate quantum dots with a controllable exchange interaction between the
two dots.  
The qubit states are represented by the two-electron spin
states with zero total spin, 
$\lvert \uparrow \rangle_1 \otimes \lvert \downarrow \rangle_2$ 
and $\lvert \downarrow \rangle_1 \otimes \lvert \uparrow \rangle_2$,
where the $1$ and $2$ subscripts label the dots (and contained
electrons).
An applied magnetic field protects each electron spin from depolarization;
at the same time, the degeneracy of the zero-spin subspace is
protected from uniform magnetic-field fluctuations.\cite{doubleQDproposal}
Electrostatic potentials are used to manipulate
the electrons.  State preparation and final readout are performed by
biasing the two electrons, with an applied voltage, 
into the same dot so that the singlet state,
$\left(\lvert \uparrow \rangle_1 \otimes \lvert \downarrow \rangle_2
+ \lvert \downarrow \rangle_1 \otimes \lvert \uparrow \rangle_2
\right) / \sqrt{2}$, has the lowest energy because of the
Pauli-exchange interaction.\cite{doubleQDproposal, doubleQD}  
Voltage control is also used to turn on an
exchange interaction by allowing the wavefunction of the two
electrons on different dots to overlap; such control can be used to
rotate the qubit.\cite{doubleQDproposal, doubleQD}  
By using this control, one can apply $\pi$ pulses in
order to perform a Hahn echo sequence or any other DD sequence 
(such as those discussed in Sec.~\ref{SecDD}) to
prolong the coherence of the qubit.

We can simply map this two-electron qubit into our single-spin qubit
formalism.  For convenience, we will define 
$\lvert 0 \rangle = 
\lvert \uparrow \rangle_1 \otimes \lvert \downarrow \rangle_2$ and
$\lvert 1 \rangle = 
\lvert \downarrow \rangle_1 \otimes \lvert \uparrow \rangle_2$ as our
two qubit basis states.
Turning on the exchange interaction will split the energies of the
$(\lvert 0 \rangle + \lvert 1 \rvert) / \sqrt{2}$ and 
$(\lvert 0 \rangle - \lvert 1 \rvert) / \sqrt{2}$ superposition states
and thereby rotate the qubit in a ``transverse'' direction as required
for a DD sequence that combats dephasing.
In order to obtain the free evolution Hamiltonian needed by our formalism, we
simply need to derive the qubit-bath Hamiltonian, $\hat{\cal H}_{eb} \hat{S}_{z}$,
from the qubit-bath interactions in each of the two dots, 
$\hat{\cal H}_{eb}^{(1)} \hat{S}_{1z}  + \hat{\cal H}_{eb}^{(2)} \hat{S}_{2z}$, 
by taking its matrix elements in terms of our
qubit basis states.  
With these definitions where we only have a dephasing coupling between
the qubit and the bath, it is clear that
$\langle 0 \vert \hat{\cal H}_{eb}^{(1)} \hat{S}_{1z}  + \hat{\cal H}_{eb}^{(2)} \hat{S}_{2z} \vert 1 \rangle = 
\langle 1 \vert \hat{\cal H}_{eb}^{(1)} \hat{S}_{1z}  + \hat{\cal H}_{eb}^{(2)} \hat{S}_{2z} \vert 0 \rangle = 0$; we thus have
only the following dephasing qubit-bath interaction:
\begin{subequations}
\begin{eqnarray}
\nonumber
\hat{\cal H}_{eb} &=& 
2 \langle 0 \vert \hat{\cal H}_{eb}^{(1)} \hat{S}_{1z}  + {\cal
  H}_{eb}^{(2)} \hat{S}_{2z} \vert 0 \rangle \\
&=& 
-2 \langle 1 \vert \hat{\cal H}_{eb}^{(1)} \hat{S}_{1z}  + {\cal
  H}_{eb}^{(2)} \hat{S}_{2z} \vert 1 \rangle \\
&=& 
\hat{\cal H}_{eb}^{(1)} - \hat{\cal H}_{eb}^{(2)}.
\end{eqnarray}
\end{subequations}
For each dot $i$, the qubit-bath interaction is given by
\begin{equation}
\label{HqbDot_i}
\hat{\cal H}_{eb}^{(i)} =
\frac{1}{2} \sum_{n \in \mbox{\scriptsize dot}~i}  A_{n}^{(i)} I_{nz} +
\frac{1}{2} \sum_{n \ne m \in \mbox{\scriptsize dot}~i} A_{nm}^{(i)} \hat{I}_{n+} \hat{I}_{m-}.
\end{equation}
During the free evolution part of the pulse sequence, the two electrons
must essentially have no overlap in their wavefunctions; therefore,
$A_n^{(i)}$ will only be nonzero when $n$ represents a nucleus in dot
$i$.  This is the justification for summing over only the relevant dot
in Eq.~(\ref{HqbDot_i}).

Assuming that the internuclear interactions occur only within the
same bath (and that the bath is initially uncorrelated), then the
problem fully decouples into spectral diffusion problems for dots $1$
and $2$ separately.  
With regards to  $\clusterCountW{1}$ in the cluster approximation 
[Eq.~(\ref{clusterApproxResult})],
we simply need to sum the cluster
contributions in the two dots separately.  
In a random unpolarized bath with two
equivalent dots, the cluster contributions in each dot will be
identical; then $v_E$ is simply
the squared value of the echo for the problem of a single electron
in just one of the dots.  There should, thus, be no qualitative
difference between the spectral diffusion of a single-spin qubit and
this double-spin qubit; a prediction of $v_E \sim
\exp{\left[-(\tau/\tau_0)^4\right]}$ for a single-spin qubit will carry over
to the double-spin qubit.

Although the reported Hahn echo decay time $T_2$ of Ref.~\onlinecite{petta05} is
compatible with our theory (which disregards other decoherence
mechanism) as a limiting case, it is clear that the experimental echo decay 
does not match the $\exp{\left[-(\tau/\tau_0)^4\right]}$ form.
The experimentalists seem to be observing a decoherence mechanism that
we are not treating.
They report that the $T_2$ time increases with an increase in magnetic
field\cite{petta05}; therefore, they must not be operating in the high
field limit regime.
Our results may be viewed as yielding the best decoherence times achievable
by increasing the applied magnetic field.

\subsection{Dynamic nuclear polarization and the ``Zamboni effect''}
\label{Zamboni}

To minimize the effects of decoherence due to a bath of
nuclear spins, one strategy is to polarize the nuclei.  
When they are polarized, they cannot flip-flop.
This is
particularly appealing in III-V semiconductors, where all of the
isotopes have nonzero spin.
Recent experiments have successfully achieved some degree of nuclear
polarization in double quantum dot
systems.\cite{doubleDotNucPolarization}
This is accomplished by biasing to a point where there is an
anticrossing between the single state and the $m_s = +1$ triplet
state; the transition between these states requires a nuclear spin
flip to conserve angular momentum.
By cycling through this anticrossing, they are able to produce
polarizations of a few percent (producing effective nuclear fields of about $20~\mbox{mT}$ in dots where full polarization would yield about $5~\mbox{T}$).\cite{doubleDotNucPolarization}

Even with such modest polarization, there can be a significant impact
on inhomogeneous broadening.  
It does this by effectively smoothing out the hyperfine field
and, because of this smoothing, has been coined the Zamboni
effect by experimentalists.\cite{doubleDotNucPolarization}
Essentially, the process of dynamic nuclear polarization will most
likely polarize those nuclei with the strongest coupling to the
electron, those with the largest hyperfine coupling.
These nuclei are also the most important in terms of inhomogeneous
broadening (they give the largest contribution to the effective
magnetic field).  
Strong polarization is not necessarily involved in the suppression of 
inhomogeneous broadening.  Homogenizing the system to remove the
broadening need only change the polarization by roughly the same amount as the
unbiased statistical broadening, which scales as $1/\sqrt{N}$; this is less than
$1\%$ for $N=10^5$.
By removing the effects of inhomogeneous broadening in this way, it
may be possible to view FID due to SD.

This modest polarization will have a weak effect on SD 
according to our theory.  While $T_2^*$ (from
inhomogeneous broadening) is improved by this strategy, $T_2$ (from
spectral diffusion) is not significantly altered.  There are two
reasons for this.  First, the nuclei being polarized are not
necessarily those nuclei responsible for SD.  This is
illustrated in Fig.~\ref{FigGaussianWF} where the regions of 
electron occupation 
probability do not correspond to the greatest SD contributors.
Second, SD has a weak dependence on polarization because its
contributors are clusters of two or more nuclei.  
Where we quantify polarization as $p = p_{\uparrow} - p_{\downarrow}$
(the difference of the probability of being up versus down assuming
spin $1/2$ nuclei), the number of pairs that can flip-flop scales as
$(1-p^2)$.\cite{DasSarmaSSC}  When the spin is larger than $1/2$, the
dependence is even weaker (there is a larger fraction of states of two
nuclei that can flip-flop).
Therefore, one needs nearly $100\%$ polarization in order to suppress
SD $T_2$ decay.

Therefore, we predict, with substantial confidence, that the coherence
enhancement by the Zamboni effect could at best lead to a decoherence
time of $T_I (\gg T_2^*)$, but never up to $T_2 (\gtrsim T_I)$,
i.e., the Zamboni effect would never produce a coherence time longer
than the spin echo coherence time.

\section{Discussion and Conclusion}
\label{conclusion}

The main result of this paper is the decoherence time formula of
Eq.~(\ref{tau0}) with wavefunction dependence 
[Eq.~(\ref{tau0gaussianFlatDot}) for the specific case of
  Gaussian-type simple harmonic oscillator confinement].  By using a
  table (Table~\ref{GaAsInAs_mu} for GaAs and InAs) 
of values for a time quantity we
  denote by $\mu$, this formula
  will yield $\tau_0^{(l)}$ for a given pulse sequence of $l$
  concatenations of the Hahn echo ($l=0$ for free induction decay, and
  $l=1$ for the Hahn spin echo itself);
  the initial behavior of the coherence or echo decay is
  $\exp{(-[\tau/\tau_0^{(l)}]^{p(l)})}$ [Eqs.~(\ref{initEchoDecay}) and
      (\ref{exponentVsConcatenations})], where $\tau$ is the time
    between pulses for the given sequence.
  By using our definitions of decoherence times, $T_I = \tau_0^{(0)}$
  (free induction decay) and $T_2 = 2 \tau_0^{(1)}$ (the traditional
  Hahn echo).  The generalized concatenated echo decoherence
  times are $T_2^{(l)} = 2^l \tau_0^{(l)}$.
An interesting experimental test of this theory could be to compare
the decoherence times of the same sample with an applied magnetic field in different
lattice directions and check for agreement with Table~\ref{GaAsInAs_mu}
(for GaAs or InAs); in such a test, however, one must carefully account for
any change in confinement as a result of changes in the applied magnetic field.
For experimental systems that allow for the application of pulse
sequences, with the ability to perform rapid $\pi$ rotations of the
electron spin relative to dynamical time scales, a more
straightforward test would be to compare different levels of
concatenation and check for agreement with Table~\ref{GaAsInAs_mu} and
Eqs.~(\ref{exponentVsConcatenations}) and (\ref{initEchoDecay}).

There are two important approximations that our decoherence time
formula assumes.  First, we take the limit of a large applied magnetic
field.  This may not always be experimentally accessible, but, in any
case, our results represent the maximum coherence times achievable by
applying a strong magnetic field.  Second, we use a short-time
approximation and discuss its validity in
Sec.~\ref{shortTimeJustification}.  Failure of the short-time
approximation does not invalidate our general cluster expansion
[Sec.~\ref{clusterApprox} and Ref.~\onlinecite{witzelHahnLong}], however;
it only means that our simple wavefunction dependent decoherence time
formula [Eq.~(\ref{tau0})] is no longer accurate.

Finally, in our considerations of the singlet-triplet (two electron)
double quantum dot scenario, we show that it is equivalent to the
single dot (one electron) case in terms of spectral diffusion assuming
negligible exchange interaction between pulses. (For a treatment that
includes the exchange interaction, see
Ref.~\onlinecite{DoubleDotWithExchange}.)  We also predict that the Zamboni
effect will have little impact upon spectral diffusion.

\section*{Acknowledgments}

We acknowledge Alexander Efros for valuable suggestions in our
preparation of this paper, Lieven Vandersypen for useful
discussions, and {\L}ukasz Cywi{\'n}ski for helpful insights.
Furthermore, we thank Dmitri Yakovlev for encouraging us to present InAs
results and Renbao Liu for assistance in obtaining charge density
estimates to obtain hyperfine strengths in InAs.
This work is supported by LPS-NSA and ARO-DTO.

\appendix
\section{Factorizability of cluster contributions}
\label{clusterFactorability}

We define $\clusterW{\cal C}$ as the contribution of cluster
${\cal C}$ to $\hat{W}$; that is, $\clusterW{\cal C}$ gives the
sum of all terms of $\clusterCountW{1}$ that involve cluster
${\cal C}$.  Thus,
\begin{equation}
\clusterCountW{1} = \sum_{{\cal C} \ne \emptyset}
\clusterW{\cal C}.
\end{equation}
Likewise, we will define $\biClusterW{\cal A}{\cal B}$ as the
sum of all terms of $\clusterCountW{2}$ that involve clusters
${\cal A}$ and ${\cal B}$ so that
\begin{equation}
\clusterCountW{2} = \frac{1}{2}
\sum_{
\substack{
{\cal A}, {\cal B} \ne \emptyset \\
{\cal A} \bigcap {\cal B} = \emptyset}}
\biClusterW{\cal A}{\cal B},
\end{equation}
where the factor of $1/2$ is necessary to compensate for the double
counting of $\biClusterW{\cal B}{\cal A} = \biClusterW{\cal A}{\cal B}$
and defining $\setW{\cal S}$ to be the solution of $\hat{W}$ when only
including nuclei in the set ${\cal S}$ (with all of the interactions
between them).
Similarly, we may define $\setW{{\cal A}, {\cal B}}$ to be the
solution of $\hat{W}$ when only including nuclei in the sets ${\cal A}$
and ${\cal B}$ with interactions among ${\cal A}$ and among ${\cal
  B}$ but not between ${\cal A}$ and ${\cal B}$.
Because $\setW{{\cal A}, {\cal B}}$ 
is just a product of evolution operators of the form
$\exp{\left(-i \left[ \hat{\cal H}_{\cal A}^{\pm} + \hat{\cal H}_{\cal
    B}^{\pm} \right] \tau\right)} = \exp{\left(-i \hat{\cal H}_{\cal
    A}^{\pm} \tau \right)} \exp{\left(-i \hat{\cal H}_{\cal
    B}^{\pm} \tau \right)}$, $\setW{{\cal A}, {\cal B}} = \setW{\cal A}
\otimes \setW{\cal B}$.
Note that $\clusterW{\cal C}$ is the ${\cal C}$ cluster
contribution to any $\setW{\cal S}$ with ${\cal S} \supset {\cal C}$;
this is simply due to the fact that any interactions of $\hat{W}$
that are not contained in $\setW{\cal S}$ are irrelevant when
considering terms that do not involve those interactions.  
For this reason, $\clusterW{\cal A}$ is the ${\cal A}$ cluster contribution
of $\setW{{\cal A}, {\cal B}} = \setW{\cal A}
\otimes \setW{\cal B}$ and $\clusterW{\cal B}$ is its ${\cal B}$
cluster contribution.
Therefore,
$\biClusterW{\cal A}{\cal B} = \clusterW{\cal A}
\otimes \clusterW{\cal B}$ so that the double cluster contribution is
simply the product of the individual cluster contributions.

This procedure may be applied to terms of any number of clusters so
that
\begin{equation}
\hat{W} = 
\sum_{
\substack{
\left\{{\cal C}_i\right\}~\mbox{\scriptsize{disjoint}}, \\
{\cal C}_i~\ne~\emptyset,
}} \prod_{\otimes~i} \clusterW{{\cal C}_i}.
\end{equation}
Assuming that the initial bath states are uncorrelated, then
$\langle \clusterW{\cal A} \otimes \clusterW{\cal B} \rangle = 
\langle \clusterW{\cal A} \rangle \times \langle \clusterW{\cal B}
\rangle$ and
\begin{equation}
\langle \hat{W} \rangle = 
\sum_{
\substack{
\left\{{\cal C}_i\right\}~\mbox{\scriptsize{disjoint}}, \\
{\cal C}_i~\ne~\emptyset,
}} \prod_i \langle \clusterW{{\cal C}_i} \rangle,
\end{equation}
which essentially reproduces our cluster decomposition from
Ref.~\onlinecite{witzelHahnLong}.

When only small clusters, relative to the size of the bath, give
non-negligible contributions to $\clusterCountW{1}$, then this
factorizability allows us to make the following approximation:
\begin{equation}
\label{clusterCountApprox}
\langle \clusterCountW{k} \rangle \approx \frac{1}{k!} \langle
\clusterCountW{1} \rangle^k,
\end{equation}
assuming $k \ll N$.
The right hand side will provide all necessary products of cluster 
contributions without overcounting (the $1/k!$ factor compensates for
permutation overcounting); however, it also includes products among
overlapping clusters.  In a large bath, sets of overlapping clusters
are negligible compared to the number of sets of nonoverlapping
clusters so that these extraneous terms are negligible.

\section{Computing the continuum approximation tensor}
\label{computingM}

Computing ${\bf M}^{(l)}$ of Eq.~(\ref{continuum2})
 by calculating $f_b^{(l)}(\rVec_m)$ 
from Eq.~(\ref{fnm}) offers the advantage of simplifications due to
the fact that operators which act on different sets of nuclei must
commute.  For example, to compute $\left[\hat{I}_{nz}, {\hat H}_{b}
    \right]$, one need only consider the terms in ${\hat H}_{b}$ that
  involve $m$.  For us, however, it was more convenient to reuse a code
  that computes $\clusterContrib{\cal C}$ for any set of nuclei,
 ${\cal C}$.
By noting Eqs.~(\ref{DeltaDagDelta}) and (\ref{factoredOutElectron}),
we may compute $f_{n, n}^{(l)}$ by letting $A_k \propto \delta_{n, k}$
and summing together the lowest-order (in the time expansion) results 
of all cluster contributions, $\clusterContrib{\cal C}$,
that include nucleus $n$ ($n \in {\cal C}$).
Similarly, if we let $A_k \propto \delta_{n, k} + \delta_{m, k}$ for a
 given pair $n \ne m$, we may compute $f_{n, n}^{(l)} + f_{m, m}^{(l)}
 + f_{n, m}^{(l)} + f_{m, n}^{(l)}$.  Subtracting off the $f_{n,
 n}^{(l)}$ and $f_{m, m}^{(l)}$ parts that may be computed by using $A_k
 \propto \delta_{n, k}$ and $A_k \propto \delta_{m, k}$, we can obtain 
$f_{n, m}^{(l)} + f_{m, n}^{(l)}$, which may then be used to compute
${\bf M}^{(l)}$ from Eqs.~(\ref{continuum2}) and (\ref{Cjl}).  
We may also use a statistical
sampling of clusters to speed up the calculation of ${\bf M}^{(l)}$,
which is particularly
useful as one increases the number of concatenations, $l$.

\end{document}